\documentclass[reprint,prd,nofootinbib,superscriptaddress]{revtex4}

\usepackage{eurosym}
\usepackage{amsfonts}
\usepackage{amsmath}
\usepackage{amssymb}
\usepackage[english]{babel}
\usepackage{graphicx}
\usepackage{epsfig}
\usepackage{bm}
\usepackage{verbatim}
\usepackage[utf8]{inputenc}
\usepackage{booktabs}
\usepackage{multirow}
\usepackage{subfig}
\usepackage{xcolor}
\usepackage[colorlinks=true,urlcolor=red,citecolor=red]{hyperref}
\usepackage[font=small]{caption}
\usepackage{float}
\usepackage{blindtext}

\newcommand{\hs}{\hspace*{0.2cm}}
\newcommand{\nn}{\nonumber}
\newcommand{\crn}{\nonumber \\}

\newcommand{\al}{\alpha}
\newcommand{\la}{\lambda}

\newcommand{\ga}{\gamma}
\newcommand{\va}{\varphi}

\newcommand{\fr}{\frac}

\newcommand{\bc}{\begin{center}}
\newcommand{\ec}{\end{center}}
\newcommand{\be}{\begin{equation}}
\newcommand{\ee}{\end{equation}}
\newcommand{\bea}{\begin{eqnarray}}
\newcommand{\eea}{\end{eqnarray}}
\newcommand{\Ga}{\Gamma}
\newcommand{\de}{\delta}
\newcommand{\De}{\Delta}
\newcommand{\ep}{\epsilon}
\newcommand{\varep}{\varepsilon}
\newcommand{\ka}{\kappa}
\newcommand{\La}{\Lambda}
\newcommand{\si}{\sigma}

\newcommand{\mathsym}[1]{}

\newcommand{\crb}[1]{{#1}}

\DeclareUnicodeCharacter{2212}{-}
\input{tcilatex}
\begin{document}

\title{Fermion masses and mixings and $g-2$ muon anomaly in a 3-3-1 model \\
with $D_4$ family symmetry}
\author{A. E. C\'arcamo Hern\'andez$^{a,b,c}$}
\email{antonio.carcamo@usm.cl}
\author{Hoang Ngoc Long$^{d}$\footnote{Corresponding author}}
\email{hnlongg@iop.vast.vn}
\author{ M. L. Mora-Urrutia$^{a}$}
\email{maria.luisa.mora.u@gmail.com}
\author{N. H. Thao$^{e}$}
\email{nguyenhuythao@hpu2.edu.vn}
\author{V. V. Vien$^{f,g}$}
\email{vovanvien@tdmu.edu.vn}
\date{\today }

\affiliation{$^a$Department of Physics, Universidad T\'{e}cnica Federico Santa Mar\'{\i}a,\\
Casilla 110-V, Valpara\'{\i}so, Chile\\
$^b$Centro Cient\'{\i}fico-Tecnol\'ogico de Valpara\'{\i}so, Casilla 110-V, Valpara\'{\i}so, Chile,\\
$^c$Millennium Institute for Subatomic physics at high energy frontier - SAPHIR, Fernandez Concha 700, Santiago, Chile\\
$^{d}$ Institute of Physics, Vietnam Academy of Science and Technology, \\
10 Dao Tan, Ba Dinh, Hanoi, Vietnam\\
$^e$Department of Physics, Hanoi Pedagogical University 2, Phuc Yen, Vinh
Phuc, Vietnam\\
$^f$Institute of Applied Technology, Thu Dau Mot University, Binh Duong Province, Vietnam \\
$^g$ Department of Physics, Tay Nguyen University, 567 Le Duan, Buon Ma
Thuot City, DakLak, Vietnam }

\begin{abstract}
We propose a predictive model based on the $SU(3)_C\times SU(3)_L\times
U(1)_X$ gauge symmetry, which is supplemented by the $D_4$ family symmetry
and several auxiliary cyclic symmetries whose spontaneous breaking produces
the observed SM fermion mass and mixing pattern. The masses of the light
active neutrinos are produced by an inverse seesaw mechanism mediated by
three right handed Majorana neutrinos. To the best of our knowledge the
model corresponds to the first implementation of the $D_4$ family symmetry
in a $SU(3)_C\times SU(3)_L\times U(1)_X$ theory with three right handed
Majorana neutrinos and inverse seesaw mechanism. Our proposed model
successfully accommodates the experimental values of the SM fermion mass and
mixing parameters, the muon anomalous magnetic moment as well as the Higgs
diphoton decay rate and meson oscillations constraints. The consistency of our model with the muon anomalous magnetic moment requires \crb{charged exotic vector like leptons at the TeV scale.}

\end{abstract}

\pacs{14.60.St, 11.30.Hv, 12.60.-i}
\maketitle

\section{\label{intro}Introduction}

Despite its great success and consistency with the experimental data,
Standard Model (SM) have several unexplained issues such as the number of SM
fermion families, the electric charge quantization, the huge SM fermion mass hierarchy, the small quark mixing
angles and the sizeable leptonic mixing ones. Whereas the quark mixing
angles are small, two of the leptonic mixing angles are large and one is of
the order of the Cabibbo angle. In addition, the SM charged fermion mass
pattern spread over a range of 13 orders of magnitude from the light active
neutrino mass scale up to the top quark mass. This is the so called flavour
puzzle of the SM which motivates the construction of several extensions of
the SM with augmented particle spectrum and extra symmetries, which be
continuous and (or) discrete, introduced to provide a successful explanation
of the observed SM fermion mass and mixing hierarchy. Discrete flavor
symmetries have been shown to be successful in describing the SM fermion
mass and mixing pattern. Some reviews of discrete flavor groups are provided
in \cite{King:2013eh,Altarelli:2010gt,Ishimori:2010au,King:2015aea}. In
particular, the $D_4$ discrete flavor group, which has a small amount of doublets and
singlets in their irreducible representations
has been employed in extensions of the SM \cite{Frampton:1994rk,Grimus:2003kq,Grimus:2004rj,Frigerio:2004jg,Blum:2007jz,Adulpravitchai:2008yp,Ishimori:2008gp,Hagedorn:2010mq,Meloni:2011cc,Vien:2013zra,Vien:2014ica,Vien:2014soa,CarcamoHernandez:2020ney,Vien:2020uzf,Vien:2021diw,Bonilla:2020hct}, since it allows to get viable
predictions for the SM fermion mass and mixing hierarchy, with a moderate
amount of particle content. Furthermore, several theories with enlarged
particle spectrum and symmetries have been constructed to explain the
experimental value of the muon anomalous magnetic moment, anomaly recently confirmed by the
muon $g-2$ experiment at FERMILAB. See ~\cite{Athron:2021iuf} for a very recent review.

To address the aforementioned issues of the SM, in this paper, we construct
a theory based on the $SU(3)_C\times SU(3)_L\times U(1)_X$ gauge symmetry
(3-3-1 model) with extended particle spectrum and discrete symmetries which
allows to get predictive SM fermion mass matrices consistent with the low
energy SM fermion flavor data. In our proposed theory, we considered the $%
SU(3)_C\times SU(3)_L\times U(1)_X$ gauge symmetry, since models having this
symmetry naturally explain the number of SM fermion families as well as the
electric charge quantization, see for instance \cite%
{Valle:1983dk,Pisano:1991ee,Frampton:1992wt,Foot:1994ym,Hoang:1995vq,CarcamoHernandez:2005ka,Chang:2006aa,Hernandez:2013mcf,Hernandez:2013hea,Boucenna:2014ela,Hernandez:2014lpa,Hernandez:2014vta,Okada:2015bxa,Hernandez:2016eod,Fonseca:2016tbn,CarcamoHernandez:2017cwi,CarcamoHernandez:2018iel,CarcamoHernandez:2019vih,CarcamoHernandez:2019iwh,CarcamoHernandez:2019lhv,CarcamoHernandez:2020pxw,CarcamoHernandez:2020ehn}%
. Apart from successfully addressing these features, our proposed model also
successfully explains and accommodates the SM fermion mass and mixing
hierarchy 
the muon anomalous magnetic moment as well as the Higgs diphoton decay rate
constraints. Our theory is based in the $D_4$ discrete symmetry, which is
supplemented by several cyclic symmetries. In our proposed theory, the SM
fermion mass and mixing pattern is produced by the spontaneous breaking of
the 
discrete symmetries, whereas the tiny masses of the light active neutrinos
are produced by an inverse seesaw mechanism mediated by three right handed
Majorana neutrinos. To the best of our knowledge our work corresponds to the
first implementation of the $D_4$ family symmetry in a $SU(3)_C\times
SU(3)_L\times U(1)_X$ theory with three right handed Majorana neutrinos and
inverse seesaw mechanism. The layout of the reminder of the paper is as
follows. In section \ref{model} we describe the proposed model. The
consequences of the model in quark masses and mixings are analyzed in
section \ref{quarks}. Lepton masses and mixings are described in section \ref%
{leptons}. The low energy scalar of the model is discussed in section \ref%
{scalars}. In section \ref{diphoton} we discuss the implications of the
model in the Higgs diphoton decay rate. The implications of the model in the
muon anomalous magnetic and meson oscillations are discussed in
sections \ref{gminus2} and \ref{FCNC}. We conclude in section \ref%
{conclusions}.

\section{The model}

\label{model} The model under consideration is based on the $SU(3)_{C}\times
SU(3)_{L}\times U(1)_{X}$ gauge symmetry, which is supplemented by the $%
D_{4}\times Z_{4}\times Z_{3}^{\left( 1\right) }\times Z_{3}^{\left(
2\right) }\times Z_{16}$ discrete group, whose spontaneous breaking
generates viable and predictive fermion mass matrices consistent with the
observed pattern of SM fermion masses and mixings. We choose the $D_{4}$
symmetry since it is the smallest non-Abelian discrete symmetry group having
five irreducible representations (irreps), explicitly, four singlets and one
doublet irreps. The auxiliary cyclic symmetries $Z_{4}$, $Z_{3}^{\left(
1\right) }$ and $Z_{3}^{\left( 2\right) }$ select the allowed entries of the
SM fermion mass matrices that yield a predictive and viable pattern of SM
fermion masses and mixings. These cyclic symmetries also allows a successful
implementation of the inverse seesaw mechanism. These symmetries together
with the $Z_{16}$ symmetry shape the hierarchical structure of the SM
charged fermion mass matrices crucial to yield the observed pattern of SM
charged fermion masses and mixing angles. Furthermore, the $Z_{16}$ discrete
symmetry is also crucial to get sufficiently suppressed non renormalizable
mass terms involving gauge singlet right handed Majorana neutrinos, required
for the implementation of the inverse seesaw mechanism that produces small
masses for the light active neutrinos. The model fermionic sector contains $%
SU(3)_{L}$ fermionic triplets and antitriplets, \crb{transforming under the
$SU(3)_{C}\times SU(3)_{L}\times U(1)_{X}$ gauge symmetry as follows:} 
\bea
&&Q_{1L}=%
\begin{pmatrix}
u_{1} \\
d_{1} \\
J_{1} \\
\end{pmatrix}%
_{L}\sim \left( \mathbf{3},\mathbf{3},\fr{1}{3}\right) ,%
\hspace{0.2cm}Q_{nL}=%
\begin{pmatrix}
d_{n} \\
-u_{n} \\
J_{n} \\
\end{pmatrix}%
_{L}\sim \left( \mathbf{3},\mathbf{\bar{3}},0\right) ,\nn
\\
&&u_{iR}\sim \left( \mathbf{3},\mathbf{1},\fr{2}{3}\right) ,%
\hspace*{0.2cm}d_{iR}\sim \left( \mathbf{3},\mathbf{1},-\fr{1}{3}\right) ,%
\hspace*{0.2cm}J_{1R}\sim \left( \mathbf{3},\mathbf{1},\fr{2}{3}\right) ,%
\hspace{0.2cm}J_{nR}\sim \left( \mathbf{3},\mathbf{1},-\fr{1}{3}\right) ,
\crn
&&L_{iL}=%
\begin{pmatrix}
\nu _{i} \\
l_{i} \\
\nu _{i}^{c} \\
\end{pmatrix}%
_{L}\sim \left( \mathbf{1},\mathbf{3},-\fr{1}{3}\right) ,%
\hspace{0.2cm}l_{iR}\sim \left( \mathbf{1},\mathbf{1},-1\right) ,%
\hspace{0.2cm}N_{iR}\sim \left( \mathbf{1},\mathbf{1},0\right) ,\hspace{%
0.2cm}n=2,3;i=1,2,3.
\eea %
All $SU(3)_{L}$ singlets $\left\{ \xi ,\hspace{0.1cm}\Xi,\hspace{0.1cm}\si ,%
\hspace{0.1cm}\phi _{1,2},\hspace{0.1cm}\Phi ,\hspace{0.1cm}\phi ,\hspace{%
0.1cm}\zeta ,\hspace{0.1cm}\eta ,\hspace{0.1cm}\va  _{1,2}\right\} $
transform as $(\mathbf{1},\mathbf{1},0)$ under the $SU(3)_{C}\times
SU(3)_{L}\times U(1)_{X}$ gauge symmetry.\newline
Furthermore, in the model fermionic sector, three right handed Majorana
neutrinos are included as well, in order to allow a successful implementation
of the inverse seesaw mechanism that produces the tiny active neutrino
masses. Notice that the fermions in our model do not feature exotic electric
charges, from which it follows that the electric charge is given by:
  \be
Q=T_{3}+\beta T_{8}+X=T_{3}-\fr{1}{\sqrt{3}}T_{8}+X.
\ee %
On the other hand, the model scalar sector is composed of two $SU(3)_{L}$
triplet scalars $\chi $ and $\rho $ and several gauge singlet scalar fields
to be specified below. The $SU(3)_{L}$ scalar  $\chi $ and $\rho $
can be expanded around the minimum as follows:
\bea
&&\chi =%
\begin{pmatrix}
\chi _{1}^{0} \\
\chi _{2}^{-} \\
\fr{1}{\sqrt{2}}(v_{\chi }+\xi _{\chi }\pm i\zeta _{\chi })%
\end{pmatrix}, \hspace{0.5cm}
\rho =\begin{pmatrix}
\rho _{1}^{+} \\
\fr{1}{\sqrt{2}}(v_{\rho }+\xi _{\rho }\pm i\zeta _{\rho }) \\
\rho _{3}^{+}%
\end{pmatrix}.
\eea
This implies that the $SU(3)_L$ scalar triplets acquire the following VEV pattern:
\bea
\crb{\langle \chi \rangle^T = \left(0\,,\hs 0 \,, \hs v_{\chi }/\sqrt{2}\right),\hspace{1cm}\langle\rho \rangle^T=\left(0\,, \hs v_{\rho }/\sqrt{2 }\, ,\hs  0 \right)\, .
}
\eea
The scalar and fermionic spectrum and their assignments under the $SU(3)_{C} \times SU(3)_{L} \times U(1)_{X} \times D_{4}\times Z_{4}\times Z_{3}^{(1) } \times Z_{3}^{(2)} \times Z_{16}$ group are shown in Tables \ref{Tabscalar} and \ref{Tabfermion}, respectively.
\begin{table}[ht]
\caption{Scalar transformations under the $SU(3)_{C} \times SU(3)_{L} \times U(1)_{X} \times D_{4}\times Z_{4}\times Z_{3}^{(1) } \times Z_{3}^{(2)} \times Z_{16}$ group.}\label{Tabscalar}
\vspace{-0.25 cm}
\begin{center}
\begin{tabular}{|c|l|l|l|l|l|l|l|l|l|l|l|l|l|l|l|l}
\hline\hline
& ~$\chi$ & ~$\rho$ & ~$\xi$ & ~$\Xi$ & ~~$\si$ & ~$\phi_{1}$ & ~$\phi _{2}$ & ~$\Phi$ & ~$\zeta$  & ~$\eta$ & ~$\va _{1}$ & ~$\va _{2}$ \\ \hline
$SU(3)_{C}$   & $\mathbf{1}$ & $\mathbf{1}$ & 1 & 1 & 1 & 1 & 1 & 1 & 1 & 1 & 1 & 1 \\ \hline
$SU(3)_{L}$   & $\mathbf{3}$ & $\mathbf{3}$ & 1 & 1 & 1 & 1 & 1 & 1 & 1 & 1 & 1 & 1 \\ \hline
$U(1)_{X}$    & $- \fr{1}{3}$ & $\fr{2}{3}$ & 0 & 0 & 0 & 0 & 0 & 0 & 0 & 0 & 0 & 0 \\ \hline
$D_{4}$       & $\mathbf{1}_{+-}$ & $\mathbf{1}_{++}$ & $\mathbf{2}$ & $\mathbf{2}$ & $\mathbf{1}_{-+}$ & $\mathbf{1}_{++}$ & $\mathbf{1}_{--}$ & $\mathbf{2}$ & $\mathbf{1}_{--}$ & $\mathbf{2}$ & $\mathbf{1}_{+-}$ & $\mathbf{1}_{++}$ \\ \hline
$Z_{4}$       & -1 & 1 & 0 & 2 & 0 & -1 & -1 & 0 & 1 & 1 & -2 & -2 \\ \hline
$Z_{3}^{(1)}$ & 0 & 0 & 0 & 0 & 0 & -1 & -1 & 0 & 0 & 0 & 0 & 0 \\ \hline
$Z_{3}^{(2)}$ & 0 & 0 & 0 & 0 & 0 & 0 & 0 & -1 & 2 & 2 & -2 & -2 \\ \hline
$Z_{16}$      & 0 & 0 & 0 & 0 & -1 & -1 & -1 & 0 & -8 & -8 & 0 & 0 \\  \hline\hline
\end{tabular}%
\end{center}
\par
\end{table}
\vspace{-0.5 cm}
\begin{table}[h]
\caption{Fermion transformations under the  $SU(3)_{C} \times SU(3)_{L} \times U(1)_{X} \times D_{4}\times Z_{4}\times Z_{3}^{(1) } \times Z_{3}^{(2)} \times Z_{16}$ group}.\label{Tabfermion}
\vspace{-0.25 cm}
\begin{center}
\begin{tabular}{|c|l|l|l|l|l|l|l|l|l|l|l|l|l|l|l|l|l|l|l|l|}
\hline\hline
& ~$Q_{1L}$ & ~$Q_{2L}$ & ~$Q_{3L}$ & ~$u_{1R}$ & ~$u_{2R}$ & ~$u_{3R}$ & ~$d_{1R}$ & ~$D_{R}$ & ~$J_{1R}$  & ~$J_{2R}$ & ~$J_{3R}$ & ~$L_{L}$ & ~$L_{3L}$ & ~$l_{1R}$ & ~$l_{2R}$ & ~$l_{3R}$ & ~$N_{R}$ & ~$N_{3R}$ & ~$E_{L}$ & ~$E_{R}$\\ \hline
$SU(3)_{C}$ & $\mathbf{3}$ & $\mathbf{3}$ & $\mathbf{3}$ & $\mathbf{3}$ & $\mathbf{3}$ & $\mathbf{3}$ & $\mathbf{3}$ & $\mathbf{3}$ & $\mathbf{3}$ & $\mathbf{3}$ & $\mathbf{3}$ & $\mathbf{1}$& $\mathbf{1}$ & $\mathbf{1}$ & $\mathbf{1}$ & $\mathbf{1}$ & $\mathbf{1}$ & $\mathbf{1}$ & $\mathbf{1}$ & $\mathbf{1}$ \\ \hline
$SU(3)_{L}$ & $\mathbf{3}$ & $\mathbf{\bar{3}}$ & $\mathbf{\bar{3}}$ & $\mathbf{1}$ & $\mathbf{1}$ & $\mathbf{1}$ & $\mathbf{1}$ & $\mathbf{1}$ & $\mathbf{1}$ & $\mathbf{1}$ & $\mathbf{1}$ & $\mathbf{3}$ & $\mathbf{3}$ & $\mathbf{1}$ & $\mathbf{1}$ & $\mathbf{1}$ & $\mathbf{1}$ & $\mathbf{1}$ & $\mathbf{1}$ & $\mathbf{1}$ \\ \hline
$U(1)_{X}$ & $\fr{1}{3}$ & 0 & 0 & $\fr{2}{3}$ & $\fr{2}{3}$ & $\fr{2}{3}$ & $- \fr{1}{3}$ & $- \fr{1}{3}$ & $\fr{2}{3}$ & $- \fr{1}{3}$ & $- \fr{1}{3}$ & $- \fr{1}{3}$ & $- \fr{1}{3}$ & $-1$ & $-1$ & $-1$ & 0 & 0 & $-1$ & $-1$ \\ \hline
    $D_{4}$ & $\mathbf{1}_{+-}$ & $\mathbf{1}_{-+}$ & $\mathbf{1}_{--}$ & $\mathbf{1}_{-+}$ & $\mathbf{1}_{-+}$ & $\mathbf{1}_{--}$ & $\mathbf{1}_{-+}$  & $\mathbf{2}$ &$\mathbf{1}_{++}$ & {$\mathbf{1}_{--}$} & {$\mathbf{1}_{-+}$} & $\mathbf{2}$ & $\mathbf{1}_{++}$ & $\mathbf{1}_{+-}$ & $\mathbf{1}_{--}$ & $  {\mathbf{1}_{-+}}$ & $\mathbf{2}$ & $  {\mathbf{1}_{+-}}$ & $\mathbf{2}$ & $\mathbf{2}$ \\ \hline
$Z_{4}$       & 0 & 0 & {2} & 0 & 1 & {-1} & -1 & 0 & 1 & -1 & {1} & 0 & 0 & -1 & -1 & -1 & 1 & 1 &  {-1} & {-1}\\ \hline
$Z_{3}^{(1)}$ & -1 & 0 & 0 & -1 & 0 & 0 & -1 & 0 & -1 & 0 & 0 & 0 & 0 & 0 & 0 & 0 & 0 & 0 & {0} & {0}\\ \hline
$Z_{3}^{(2)}$ & 0 & 0 & 0 & 0 & 0 & 0 & 0 & 0 & 0 & 0 & 0 & 1 & 1 & {-1} & {-1} & {1} & 1 & 1 & {1} & {1}\\ \hline
$Z_{16}$      & -4 & -2 & 0 & {3} & 2 & 0 & 0 & 1 & -4 & -2 & 0 & -4 & -4 & {0} & {-4} & {-1} & {-4} & {-4} & {-4} & {-4} \\  \hline\hline
\end{tabular}%
\end{center}
\par
\end{table}

With the fermion and scalar contents in {Tables \ref{Tabscalar} and \ref{Tabfermion}}, 
the following quark and lepton Yukawa terms arise:
\bea
-\mathcal{L}_{Y}^{\left( q\right) } &=&y_{1}^{\left( J\right) }\overline{Q}%
_{1L}\chi J_{1R}+\sum_{n=2}^{3}y_{n}^{\left( J\right) }\overline{Q}_{nL}\chi
^{\ast }J_{nR}+y_{33}^{\left( u\right) }\overline{Q}_{3L}\rho ^{\ast
}u_{3R}+y_{22}^{\left( u\right) }\overline{Q}_{2L}\rho ^{\ast }u_{2R}\fr{%
\si ^{4}}{\La ^{4}}+y_{11}^{\left( u\right) }\varep _{abc}\overline{Q}_{1L}^{a}\left( \rho ^{\ast }\right) ^{b}\left( \chi ^{\ast }\right)^{c}u_{1R}  {\fr{\si ^{7}}{\La^{7}} }\crn
&+&y_{11}^{\left( d\right) }\overline{Q}_{1L}\rho d_{1R}\fr{\si
^{4}\left( \xi \xi \right) _{\mathbf{1}_{-+}}\left( \xi \xi \right) _{%
\mathbf{1}_{+-}}}{\La ^{8}}+y_{12}^{\left( d\right) }\overline{Q}%
_{1L}\rho \left( \xi D_{R}\right) _{\mathbf{1}_{  {+-}}}\fr{\phi _{1}\si
^{4}}{\La ^{6}}\,+ {\,y_{13}^{\left( d\right) }\overline{Q}_{1L}\rho
\left( \xi D_{R}\right) _{\mathbf{1}_{-+}}\fr{\phi _{2}\si ^{4}}{%
\La ^{6}}}  \crn
&+&y_{22}^{\left( d\right) }\varep _{abc}\overline{Q}_{2L}^{a}\rho
^{b}\chi ^{c}\left( \xi D_{R}\right) _{\mathbf{1}_{+-}} {\fr{\si ^{3}}{\La^{5}}}+ {y_{33}^{\left( d\right) }\varep _{abc}\overline{Q}%
_{3L}^{a}\rho ^{b}\chi ^{c}\left( \Xi D_{R}\right) _{\mathbf{1}_{++}}\fr{\si}{\La ^3}}\,+\,H.c,  \label{Lyq} \\
-\mathcal{L}_{Y}^{\left( l\right) } &=&  {y_{1}^{\left( l\right) }\left(\overline{L}_{L}\rho E_R\right) _{\mathbf{1}_{++}}+z_{1}^{\left( l\right) }\left(\overline{E}_{L}\Phi\right)_{\mathbf{1}_{+-}}l_{1R}\fr{\si^4}{\La^4}+z_{2}^{\left( l\right) }\left(\overline{E}_{L}\Phi\right)_{\mathbf{1}_{--}}l_{2R}+m_E\left(\overline{E}_{L}E_R\right)_{\mathbf{1}_{++}}
+y_{3}^{\left( l\right) }\overline{L}_{3L}\rho l_{3R}\fr{\si^3}{\La ^{3}}}  \crn
&+&  {y_{1\chi }^{\left( L\right) }\left( \overline{L}_{L}\chi \right) _{\mathbf{2}}N_{R}+y_{2\chi }^{\left( L\right) }\overline{L}_{3L}\chi N_{3R}}+x_{\rho }^{\left( 1\right) }\varep _{abc} {\left[\left(\overline{L^C_{L}}\right)^{a}\left( L_{L}\right)^{b}\right]_{\mathbf{1}_{--}}\left(\rho\right)^{c}\fr{\zeta^*}{\La }}  \crn
&+&x_{\rho }^{\left( 2\right) }\varep _{abc} {\left[\left(\overline{L^C_{L}}\right)^{a}\left( L_{3L}\right)^{b} \left(\rho\right)^{c}\right]_{\mathbf{2}}\fr{\eta^*}{\La }}+x_{\rho }^{\left( 3\right)}\varep _{abc}  {\left[\left(\overline{L^C_{3L}}\right)^{a}\left( L_{L}\right)^{b} \left(\rho\right)^{c}\right]_{\mathbf{2}}\fr{\eta^*}{\La }}  \crn
&+&y_{N_{1}}\left(  {\overline{N_{R}^{C}}N_{R}}\right) _{\mathbf{1}%
_{+-}}\va  _{1}\fr{\si ^{8}}{\La ^{8}}+y_{N_{2}}\left(  {\overline{N_{R}^{C}}N_{R}}\right) _{\mathbf{1}_{++}}\va  _{2}\fr{\si ^{8}}{%
\La ^{8}}+y_{N_{3}} {\overline{N_{3R}^{C}}N_{3R}}\va  _{2}\fr{\si
^{8}}{\La ^{8}}+H.c.  \label{Lyl}
\eea

The large amount of parametric freedom of the scalar potential allows to
consider the following vacuum expectation value (VEV) configuration for the $%
D_4$ doublets SM gauge singlet scalars:
\bea
&&  {\langle \xi \rangle =\left(v_{\xi_1},v_{\xi_2}\right)=v_{\xi}\left(1,r\right), \hspace{0.2cm}\langle\Xi \rangle =v_{\Xi}\left(1,0\right), \hspace{0.2cm}\left\langle \eta
\right\rangle =\left(v_{{\eta_1}}, v_{{\eta_2}}\right)},\hspace{0.2 cm}\langle\Phi\rangle =(v_1, v_2),\hspace{0.2cm}
\eea
  {whereas for the VEVs of the gauge singlet scalars one has:}
\be
\langle\si \rangle =v_\si ,\hspace{0.2cm}\langle \phi_{1}\rangle
=v_{\phi_{1}},\hspace{0.2cm} \langle\phi _{2}\rangle = v_{\phi_{2}},\hspace{0.2cm}\langle\phi\rangle =v_\phi,  \hspace{0.2cm}\langle\zeta\rangle =v_\zeta,\hspace{0.2 cm}
\langle\va _{1}\rangle =v_{\va _{1}},\hspace{0.2 cm}
\langle\va _{2}\rangle =v_{\va _{2}}.  \label{vevscalar}
\ee
The above given VEV pattern allows to get a predictive and viable pattern of SM fermion masses and mixings as it will be shown in the next sections.

\section{Quark masses and mixings}

\label{quarks} From the quark Yukawa terms given in Eq.$(\ref{Lyq})$, it
follows that the SM mass matrices for quarks are:
\bea
M_{U}&=&\left(
\begin{array}{ccc}
a_{1}^{(U)} \la  ^{  {8}} & 0 & 0 \\
0 & a_{2}^{(U)} \la  ^{4} & 0 \\
0 & 0 & a_{3}^{(U)} %
\end{array}%
\right),\notag\\
M_{D}&=&\left(
\begin{array}{ccc}
a_{  {1}}^{(D)} \la  ^{  {8}} & {\left(a_{4}^{(D)}+a_{5}^{(D)}\right) \la  ^{6}} & {\left(a_{4}^{(D)}-a_{5}^{(D)}\right)  r \la ^{6}} \\
0 & a_{  {2}}^{(D)} \la  ^{  {5}} & {a_{23}^{(D)} \la  ^{5}} \\
0 & 0 & a_{  {3}}^{(D)} \la  ^{  {3}}%
\end{array}%
\right)=\left(
\begin{array}{ccc}
a_{  {1}}^{(D)}\la  ^{  {8}} & {a_{12}^{(D)}\la  ^{6}} & {a_{13}^{(D)} \la ^{6}} \\
0 & a_{  {2}}^{(D)}\la  ^{  {5}} & {a_{23}^{(D)}\la  ^{5}} \\
0 & 0 & a_{  {3}}^{(D)}\la  ^{  {3}}%
\end{array}%
\right),  \label{MuMd}
\eea
where $a_{1}^{(U)}, a_{  {1}}^{(D)},...$ are $\mathcal{O}(1)$ dimensionless
parameters {which are given by:}
  {
\bea
&&a_1^{(U)}= \fr{y_{11}^{(u)}}{2}v_\rho,\hspace{0.2cm} {a_2^{(U)}=\fr{y_{22}^{(u)}}{\sqrt{2}} v_\rho, \hspace{0.2 cm} a_3^{(U)}=\fr{y_{33}^{(u)}}{\sqrt{2}} v_\rho},\crn 
&&a_1^{(D)}= \fr{y_{11}^{(d)}}{\sqrt{2}} v_\rho,  \hspace{0.2 cm} a_{2}^{(D)}=\fr{y_{22}^{(d)}}{2} v_\rho, \hspace{0.2 cm} a_{3}^{(D)}=\fr{y_{33}^{(d)}}{2} v_\rho,  \hspace{0.2 cm} a_{4}^{(D)}=\fr{y_{12}^{(d)}}{\sqrt{2}} v_\rho, \hspace{0.2 cm} a_{5}^{(D)}=\fr{y_{13}^{(d)}}{\sqrt{2}} v_\rho, \crn
&&  {\fr{v_\Xi}{\La}}\sim\fr{v_\xi}{\La}\sim \fr{v_\chi}{\La}\sim \fr{v_{\phi_1}}{\La}\sim \fr{v_{\phi_2}}{\La}\sim \fr{v_\si}{\La}=\la .
\eea}
Here $v=246$ GeV is the scale of electroweak symmetry breaking and the Wolfenstein parameter $\la  =0.225$ is used for
characterization of the hierarchy between the parameters defining quark mass
matrix elements in Eq.(\ref{MuMd}). We find that the experimental values
for the physical quark mass spectrum \cite{Xing:2019vks,Zyla:2020zbs},
mixing angles and CP violating phase~\cite{Xing:2019vks,Zyla:2020zbs} can be
well reproduced for the following benchmark point:

\bea
a_{1}^{\left(U\right) } &\simeq &1.085,\hspace{1cm}               a_{2}^{\left(U\right)}\simeq 1.413,\hspace{1cm}                   a_{3}^{\left(U\right)}\simeq 0.994,\hspace{1cm}       a_{1}^{\left(D\right) }\simeq 2.329, \hspace{1cm}            a_{2}^{\left(D\right) } \simeq 0.554, \hspace{1cm}\crn
a_{3}^{\left(D\right)} &\simeq& 1.439,\hspace{1cm}               a_{12}^{\left(D\right) }\simeq 0.570,\hspace{1cm}
a_{13}^{\left(D\right) }\simeq -(0.123 + 0.438i),\hspace{1cm}
a_{23}^{\left(D\right) }\simeq 1.152.                       \label{fit-q}
\eea

\begin{table}[th]
\caption{Model and experimental values of the quark masses and CKM
parameters.}
\label{Tab}
\begin{center}
\begin{tabular}{c|l|l}
\hline\hline
Observable   & Model value   & Experimental value \\ \hline
$m_{u}[MeV]$ & \quad $1.24$  & \quad $1.24 \pm 0.22$ \\ \hline
$m_{c}[GeV]$ & \quad $0.63$   & \quad $0.63 \pm 0.02$ \\ \hline
$m_{t}[GeV]$ & \quad $172.9$ & \quad $172.9 \pm 0.4$ \\ \hline
$m_{d}[MeV]$ & \quad $2.59$   & \quad $2.69 \pm 0.19$ \\ \hline
$m_{s}[MeV]$ & \quad $57.0$  & \quad $53.5 \pm 4.6$ \\ \hline
$m_{b}[GeV]$ & \quad $2.86$  & \quad $2.86 \pm 0.03$ \\ \hline
$\sin \theta _{12}$ & \quad $0.226$   & \quad $0.22650 \pm 0.00048$ \\ \hline
$\sin \theta _{23}$ & \quad $0.0405$  & \quad $0.04053^{+0.00083}_{-0.00061}$ \\ \hline
$\sin \theta _{13}$ & \quad $0.00360$ & \quad $0.00361^{+0.00009}_{-0.00011}$ \\ \hline
$J_{q}$ & \quad $0.0000309$ & \quad $ \left(3.00^{+0.15}_{-0.09} \right) \times 10^{-5}$ \\ \hline\hline
\end{tabular}%
\end{center}
\par
\end{table}
\vspace{3mm}

\begin{center}
\begin{figure}[h]
\begin{center}
\includegraphics[scale=0.5]{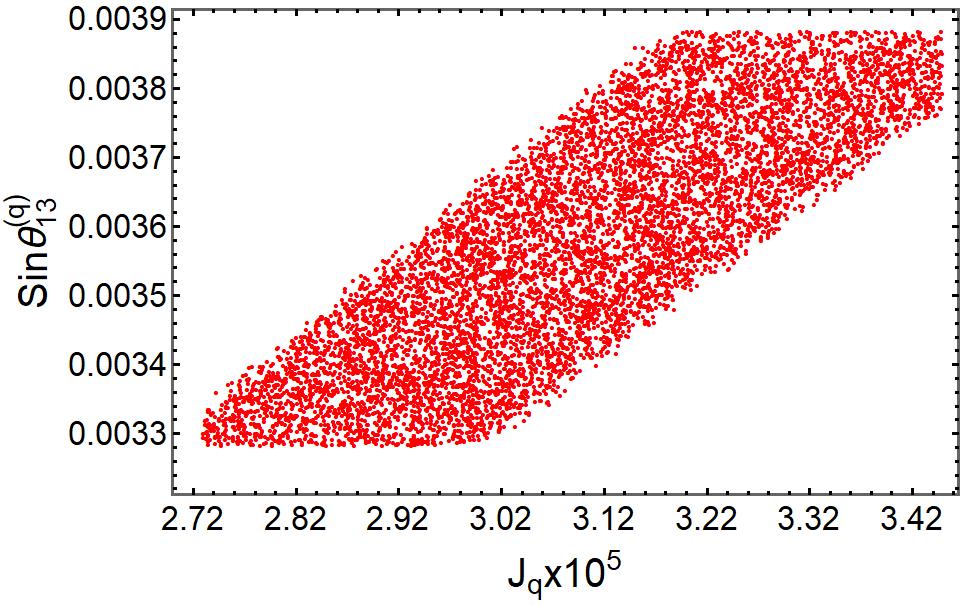}
\end{center}
\caption{Correlation plot between the quark mixing parameter $\sin\protect%
\theta_{13}$ and the Jarlskog invariant.}
\label{fig:correlplot}
\end{figure}
\end{center}

In addition, Fig.\ref{fig:correlplot} shows the correlation plot between the
quark mixing parameter $\sin\theta_{13}$ and the Jarlskog invariant. As
indicated by Fig.\ref{fig:correlplot}, $\sin\theta_{13}$ is predicted to be
in range $0.0033 \lesssim\sin\theta_{13} \lesssim 0.0040$ in the allowed
parameter space. Furthermore, the quark mixing parameter $\sin\theta_{13}$
increases when the Jarlskog invariant takes larger values.

\vspace{1mm}


\section{Lepton masses and mixings}
\label{leptons} From the charged-lepton Yukawa interactions in Eq.(\ref{Lyl}%
) and the VEV alignments in Eq. (\ref{vevscalar}), we find the following
charged -lepton mass terms: {
\bea
-\mathcal{L}^{l}_{Y}&=& (\bar{l}_{1L} \hspace*{0.2cm} \bar{l}_{2L} \hspace*{%
0.2cm} \bar{l}_{3L} \hspace*{%
0.2cm} \bar{E}_{1L} \hspace*{%
0.2cm} \bar{E}_{2L}) M_l \left(
\begin{array}{c}
l_{1R} \\
l_{2R} \\
l_{3R} \\
E_{1R} \\
E_{2R} \\
\end{array}%
\right)+ \mathrm{H.c,}  \label{Lclep}
\eea
where the charged-lepton mass matrix is given by
\bea
M_{E} &=&\left(
\begin{array}{cc}
C_{E} & A_{E} \\
B_{E} & X_{E}%
\end{array}%
\right) ,\hspace{0.7cm}\hspace{0.5cm}C_{E}=\left(
\begin{array}{ccc}
0 & 0 & 0 \\
0 & 0 & 0 \\
0 & 0 & y_{3}^{\left( l\right) }%
\end{array}%
\right) \fr{v_{\si }^{3}v_{\rho }}{\La ^{3}\sqrt{2}}=\left(
\begin{array}{ccc}
0 & 0 & 0 \\
0 & 0 & 0 \\
0 & 0 & c%
\end{array}%
\right) \fr{v_{\rho }}{\sqrt{2}}, \\
A_{E} &=&y_{1}^{\left( l\right) }\left(
\begin{array}{cc}
0 & 1 \\
1 & 0 \\
0 & 0%
\end{array}%
\right) \fr{v_{\rho }}{\sqrt{2}},\hspace{0.7cm}B_{E}=\left(
\begin{array}{ccc}
z_{1}^{\left( l\right) }\fr{v_{\si }^{2}}{\La ^{2}}v_{1} &
z_{2}^{\left( l\right) }v_{2} & 0 \\
z_{1}^{\left( l\right) }\fr{v_{\si }^{2}}{\La ^{2}}v_{2} &
-z_{2}^{\left( l\right) }v_{1} & 0%
\end{array}%
\right) =\left(
\begin{array}{ccc}
av_{1} & bv_{2} & 0 \\
av_{2} & -bv_{1} & 0%
\end{array}%
\right) , \\
X_{E} &=&\left(
\begin{array}{cc}
0 & 1 \\
1 & 0%
\end{array}%
\right) m_{E},\hspace{0.7cm}\hspace{0.7cm}a=z_{1}^{\left( l\right) }\fr{%
v_{\si }^{2}}{\La ^{2}},\hspace{0.7cm}\hspace{0.7cm}b=z_{2}^{\left(
l\right) },\hspace{0.7cm}\hspace{0.7cm}c=y_{3}^{\left( l\right) }\fr{%
v_{\si }^{3}}{\La ^{3}}.
\eea
Then, the SM charged lepton mass matrix takes the form:
  \be
M_{l}=C_{E}+A_{E}X_{E}^{-1}B_{E}=\left(
\begin{array}{ccc}
\al v_{1} & \beta v_{2} & 0 \\
\al v_{2} & -\beta v_{1} & 0 \\
0 & 0 & c\fr{v_{\rho }}{\sqrt{2}}%
\end{array}%
\right)\equiv \left(%
\begin{array}{ccc}
m_{11} & -m_{12} & 0 \\
m_{21} & m_{22} & 0 \\
0 & 0 & m_{33} \\
&  &
\end{array}%
\right)
\ee }
Let us define a Hermitian matrix $M_l$ as follows
\bea
m^2_l&=& M_{l} M^+_{l} = \left(%
\begin{array}{ccc}
|m_{11}|^2 +|m_{12}|^2 & \hspace*{0.2cm} m_{11} m^2_{21} -m_{12} m^*_{22} & 0
\\
(m_{11} m^*_{21} -m_{12} m^*_{22})^* & \hspace*{0.2cm} |m_{21}|^2+ |m_{22}|^2
& 0 \\
0 &  & |m_{33}|^2 \\
&  &
\end{array}%
\right),  \label{ml}
\eea
which can be diagonalized by $U_{L, R}$ satisfying $U^+_{L} m^2_l \,U_R=%
\mathrm{diag} (m^2_e, m^2_\mu, m^2_\tau)$, where
\bea
&& U_L =U_R=\left(%
\begin{array}{ccc}
\mathrm{cos} \theta & -\mathrm{sin} \theta . e^{-i \al} & 0 \\
\mathrm{sin} \theta . e^{i \al} & \mathrm{cos} \theta & \,\,\,\,\,\, \\
0 & \hspace*{0.2cm} & 1 \\
&  &
\end{array}%
\right),  \label{Uclepal} \\
&& m^2_{e, \mu}= \la _1 \mp \la _2,\hspace*{0.2cm} m^2_\tau
=|m_{33}|^2,  \label{memt}
\eea
with
\bea
2\la _2&=&\left\{|m_{11}|^4 + (|m_{21}|^2 - |m_{12}|^2)^2 + 2 (|m_{21}|^2
+ |m_{12}|^2) |m_{22}|^2 + 2 |m_{11}|^2 (|m_{21}|^2 + |m_{12}|^2 -
|m_{22}|^2) \right.  \crn
&+&\left. |m_{22}|^4 -8 |m_{11}| |m_{21}| |m_{12}| |m_{22}| \cos
(\de _{12}-\beta_{12})\right\}^{\fr{1}{2}}, \hspace*{0.2cm}
\de _{12}=\al_1-\al_2, \beta_{12}=\beta_1-\beta_2,  \label{la2} \\
2\la _1&=& |m_{11}|^2 + |m_{21}|^2 + |m_{12}|^2 + |m_{22}|^2, \hspace*{%
0.2cm} \al_i = \arg(a_i),\hspace*{0.2cm} \beta_i = \arg(b_i) \hspace*{%
0.2cm} (i=1,2),  \label{la1} \\
\al &=&\fr{i}{2}\log \left(\fr{m_{11} m^*_{21} - m_{12} m^*_{22}}{%
m^*_{11} m_{21} - m^*_{12} m_{22}}\right), \hspace*{0.2cm}
\theta=\arctan\left(\fr{(m^*_{11}m_{21} - m^*_{12} m_{22}) e^{-i \al}}{%
|m_{11}|^2 + |m_{12}|^2 - m^2_\mu}\right).  \label{altheta}
\eea
By comparing the obtained result in Eq. (\ref{memt}) with the experimental
values of the charged-lepton masses taken from Ref.\cite{Zyla:2020zbs}, $%
m_e=0.51099 \,\mathrm{MeV}, m_\mu = 105.65837\,\mathrm{MeV}, m_\tau =
1776.86 \,\mathrm{MeV}$, we obtain:
\bea
|m_{33}| = 1.77686 \times 10^{9}\, \mathrm{eV}, \la _1=5.58198 \times
10^{15}\, \mathrm{eV}^2,\, \la _2=5.58172 \times 10^{15} \, \mathrm{eV}^2.
\eea
In the case $v_1=v.e^{i\vartheta_1}, \, v_2=v.e^{i\vartheta_2}$, we get:
\bea
|m_{11}|&=&|m_{21}|=3.61324\times 10^5 \, \mathrm{eV}, \hspace*{0.2cm}
|m_{12}|=|m_{22}|=7.47117\times 10^7 \, \mathrm{eV}.  \label{mijabsolute}
\eea
As we will see below, since the charged lepton mixing matrix $U_L$ is non
trivial, it can contribute to the leptonic mixing matrix, defined by $%
U=U^+_{L} U_{\nu}$ where $U_{\nu}$ being neutrino mixing matrix.

\begin{figure}[tbp]
\centering
\includegraphics[width = 0.9\textwidth]{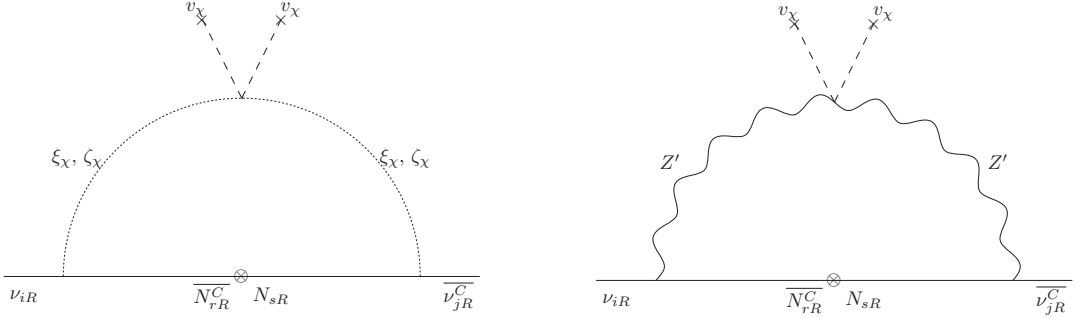}\vspace{-15cm}
\caption{Feynman diagram contributing to the 22 block of the full neutrino mass matrix. Here, $i,j,r,s=1,2,3$ and the cross mark $\otimes$ in the internal lines corresponds to the Majorana mass terms induced by the dimension twelve Majorana neutrino Yukawa interactions of Eq. (\ref{Lyl}).}
\label{Loopdiagrammutilde}
\end{figure}

Regarding the neutrino sector, from the lepton Yukawa terms in Eq. (\ref{Lyl}%
) and the VEV alignments in Eq. (\ref{vevscalar}), we find the following
neutrino mass terms:

  \be
-\mathcal{L}_{mass}^{\nu}=\fr{1}{2}\left(
\begin{array}{ccc}
\overline{\nu _{L}^{C}} & \overline{\nu _{R}^{  {C}}} & \overline{N_{R}^{  {C}}}%
\end{array}%
\right) M_{\nu }\left(
\begin{array}{c}
\nu _{L} \\
\nu _{R} \\
N_{R}\\
\end{array}%
\right) +H.c,  \label{Lnu}
\ee %
where the neutrino mass matrix reads:

  \be
M_{\nu }=%
\begin{pmatrix}
0_{3\times 3} & M_{\nu _{D}} & 0_{3\times 3} \\
M_{\nu _{D}}^{T} & 0_{3\times 3} & M_{\chi } \\
0_{3\times 3} & M_{\chi }^{T} & M_{R}%
\end{pmatrix}%
,
\ee %
and the submatrices are given by:
\bea
M_{\nu _{D}} &=& \left(
\begin{array}{ccc}
0 & -a & - b_1 \\
a & 0 & - b_2 \\
b_1 & b_2 & 0%
\end{array}%
\right)\fr{1}{\La }\fr{v_{\rho }}{\sqrt{2}},\hspace{0.25cm} M_{\chi
} {\,=\,} 
 {m_{N}}\left(
\begin{array}{ccc}
  {1} & {0} & 0 \\
  {0} & {1} & 0 \\
0 & 0 & x%
\end{array}%
\right) ,  \hspace{0.25cm}
M_{R} {\,=\,}
\left(
\begin{array}{ccc}
\ka_{1} & \ka_{2} & 0 \\
\ka_{2} & \ka_{1} & 0 \\
0 & 0 & \ka_{3}%
\end{array}%
\right) \mu ,  \label{MR}
\eea
with
\bea
  {a}&  {=}&  {x^{(1)}_\rho v_\zeta, \hs b_{1,2}=x^{(2)}_\rho v_{\eta_{2,1}}=-x^{(3)}_\rho v_{\eta_{2,1}},} \crn
x &=&\fr{y_{2\chi }^{\left( L\right) }}{y_{1\chi }^{\left( L\right) }},
\hspace{0.2cm} m_{N}=y_{1\chi }^{\left( L\right) }\fr{v_{\chi }}{\sqrt{2}}%
, \hspace{0.2cm} \ka_{1} =y_{1N}, \hspace{0.2cm} \ka_{2}=y_{2N}\fr{%
v_{\va  _{2}}}{v_{\va  _{1}}},\hspace{0.2cm}\ka_{3}=y_{3N}\fr{%
v_{\va  _{2}}}{v_{\va  _{1}}}, \hspace{0.2cm} \mu =\fr{v_{\va
_{1}}v_{\si }^{8}}{\La ^{8}}.
\eea
  {It is worth mentioning that the 22 block of the full neutrino mass matrix can be generated from the Feynman
diagram of Figure \ref{Loopdiagrammutilde}, which involves the virtual exchange of $\xi_{\chi}$, $\zeta_{\chi}$, $Z^{\prime }$ as well as the Majorana mass terms in the internal lines of the loop. These Majorana mass terms arise from the non renormalizable Majorana neutrino Yukawa interactions of Eq. (\ref{Lyl}). 
Given that the non renormalizable Majorana neutrino Yukawa terms are of dimension 12 as shown in Eq. (\ref{Lyl}), we have that the entries of the the 22 block of the full neutrino mass matrix are much smaller than the entries  of the $M_R$ submatrix and thus they give very subleading corrections to the physical neutrino mass matrices.}

The light active masses arise from an inverse seesaw mechanism and the physical neutrino mass matrices are:
\bea
M_{\nu }^{\left( 1\right) } &=&M_{\nu _{D}}\left( M_{\chi }^{T}\right)
^{-1}M_{R}M_{\chi }^{-1}M_{\nu _{D}}^{T},  \label{Mnu1} \\
M_{\nu }^{\left( 2\right) } &=&-\fr{1}{2}\left( M_{\chi }+M_{\chi
}^{T}\right) +\fr{1}{2}M_{R},\hspace{0.5cm} M_{\nu }^{\left( 3\right) }=%
\fr{1}{2}\left( M_{\chi }+M_{\chi }^{T}\right) +\fr{1}{2}M_{R},
\label{Mnu2}
\eea %
where $M_{\nu }^{\left( 1\right) }$ is the light active neutrino mass matrix
whereas $M_{\nu }^{\left( 2\right) }$ and $M_{\nu }^{\left( 3\right) }$ are
the exotic Dirac neutrino mass matrices. It is worth mentioning that
physical neutrino spectrum consists of three light active neutrinos and six
exotic neutrinos. The exotic neutrinos are pseudo-Dirac, with masses $\sim
\pm v_{\chi }\sim \mathcal{O}(10)$ TeV and a small splitting $\sim \mu $.

The mass matrix for light active neutrinos takes the form:
  \be
M_{\nu }^{\left( 1\right) }=\left(
\begin{array}{ccc}
a^2 \ka_1 + \fr{b_1^2 \ka_3}{x^2} & - a^2 \ka_2 + \fr{b_1
b_2 \ka_3}{x^2} & -a (b_2 \ka_1 + b_1 \ka_2) \\
- a^2 \ka_2 + \fr{b_1 b_2 \ka_3}{x^2} & a^2 \ka_1 + \fr{%
b_2^2 \ka_3}{x^2} & a (b_1 \ka_1 + b_2 \ka_2) \\
-a (b_2 \ka_1 + b_1 \ka_2) & a (b_1 \ka_1 + b_2 \ka_2) &
(b_1^2 + b_2^2) \ka_1 + 2 b_1 b_2 \ka_2 \\
\end{array}%
\right) m_{\nu },
\ee
where:
  \be
m_{\nu }=\fr{v_{\rho }^{2}}{2\La ^{2}}m_{N}^{2}\mu .
\ee
The mass matrix $M_{\nu }^{( 1) }$ has three exact eigenvalues as follows:
\bea
m_1&=&0,\hspace*{0.2cm} m_{2,3}=\Bbbk_1 \mp \Bbbk_2,  \label{m123}
\eea
where
\bea
\Bbbk_1&=&\fr{m_\nu}{2}\left[2 a^2 \ka_1 + (b_1^2 + b_2^2) \ka_1 +
2 b_1 b_2 \ka_2 + \fr{(b_1^2 + b_2^2) \ka_3}{x^2}\right], \hspace*{%
0.2cm} \Bbbk_2=\fr{m_\nu}{2}\fr{\sqrt{\Bbbk}}{x^2},  \label{kapal2k} \\
\Bbbk &=& (b_1^2 + b_2^2)^2 \ka_3^2 - 2 \left[(b_1^2 + b_2^2)^2 \ka
_1 +2 b_1 b_2 (2 a^2 + b_1^2 + b_2^2) \ka_2\right] \ka_3 x^2\nn
\\
&+& \left[(b_1^2 + b_2^2)^2 \ka_1^2 + 4 b_1 b_2 (2 a^2 + b_1^2 + b_2^2)
\ka_1 \ka_2 +4 (a^2 + b_1^2) (a^2 + b_2^2) \ka_2^2\right] x^4,
\label{ka}
\eea
and the corresponding mixing matrix is:
\bea
R=\left(
\begin{array}{ccc}
\fr{K_1}{\sqrt{K_1^2+N_1^2+1}} & \fr{K_2}{\sqrt{K_2^2+N_2^2+1}} & \fr{%
K_3}{\sqrt{K_3^2+N_3^2+1}} \\
\fr{N_1}{\sqrt{K_1^2+N_1^2+1}} & \fr{N_2}{\sqrt{K_2^2+N_2^2+1}} & \fr{%
N_3}{\sqrt{K_3^2+N_3^2+1}} \\
\fr{1}{\sqrt{K_1^2+N_1^2+1}} & \fr{1}{\sqrt{K_2^2+N_2^2+1}} & \fr{1}{%
\sqrt{K_3^2+N_3^2+1}}%
\end{array}%
\right)P,  \label{neumix}
\eea
where $P=\mathrm{diag}(1,\,1,\,i)$ and $K_{1,2,3},\, N_{1,2,3}$ are defined
as
\bea
&& K_1 = \fr{b_2}{a},\hspace*{0.2cm} N_1=-\fr{b_{2}}{a},\hspace*{0.2cm}
K_{2}=\ka_{21} +\ka_{22},\hspace*{0.2cm} K_{3}=\ka_{31}-\ka
_{22}, \hspace*{0.2cm} N_{2} =\ep_{21} + \ep_{22}, \hspace*{0.2cm}
N_{3} =\ep_{31} - \ep_{22},  \label{KN123}
\eea
where
\bea
\ka_{21}&=&\left[2 (a^2 + b_1^2) b_2 \ka_2 x^2 + b_1^3 (\ka
_1x^2-\ka_3)\right]\ka_0, \hspace*{0.2cm} \ka_{22}= b_1 \left[%
b_2^2 (\ka_1 x^2-\ka_3) + \sqrt{\Bbbk}\right]\ka_0,  \crn
\ka_{31}&=&\ka_{21}+\left[b_1 b_2^2 (\ka_1 x^2-\ka_3)\right]%
\ka_0, \hspace*{0.2cm} \ep_{21}=\left[2 b_1 (a^2 + b_2^2) \ka_2
x^2 + b_2(b_1^2+ b_2^3)(\ka_1 x^2-\ka_3)\right]\ka_0,  \crn
\ep_{22} &=& b_2 \sqrt{\Bbbk}\ka_0,\hspace*{0.2cm} \ep_{31}
=\left\{\left[b_2 (b_1^2 + b_2^2) \ka_1 + 2 b_1 (a^2 + b_2^2) \ka_2%
\right] x^2-b_2 (b_1^2 + b_2^2) \ka_3\right\}\ka_0,  \crn
\ka_0&=& \left[2 a (b_1^2 - b_2^2)\ka_2 x^2\right]^{-1}.
\label{k11kl2ep11ep12}
\eea
It is easy to check that $K_{i}, N_{i}\, (i=1,2,3)$ satisfy the following
relations
\bea
&& 1 + K_1 K_2 + N_1 N_2=0,\,\, 1 + K_1 K_3 + N_1 N_3=0,\,\, 1 + K_2 K_3 +
N_2 N_3=0.  \label{relation}
\eea
The matrix $M_{\nu }^{\left( 1\right) }$ is diagonalized as
  \be
U_{\nu }^\mathrm{\dagger} M_{\nu }^{\left( 1\right)} U_{\nu }=\left\{
\begin{array}{l}
\left(
\begin{array}{ccc}
0 & 0 & 0 \\
0 & m_{2} & 0 \\
0 & 0 & m_{3}%
\end{array}%
\right) ,\hspace{0.1cm} U_{\nu }=\left(
\begin{array}{ccc}
\fr{K_1}{\sqrt{K_1^2+N_1^2+1}} & \fr{K_2}{\sqrt{K_2^2+N_2^2+1}} & \fr{%
iK_3}{\sqrt{K_3^2+N_3^2+1}} \\
\fr{N_1}{\sqrt{K_1^2+N_1^2+1}} & \fr{N_2}{\sqrt{K_2^2+N_2^2+1}} & \fr{%
iN_3}{\sqrt{K_3^2+N_3^2+1}} \\
\fr{1}{\sqrt{K_1^2+N_1^2+1}} & \fr{1}{\sqrt{K_2^2+N_2^2+1}} & \fr{i}{%
\sqrt{K_3^2+N_3^2+1}}%
\end{array}%
\right) \hspace{0.1cm}\mbox{for NH,}\ \  \\
\left(
\begin{array}{ccc}
m_{3} & 0 & 0 \\
0 & m_{2} & 0 \\
0 & 0 & 0%
\end{array}%
\right) ,\hspace{0.1cm} U_{\nu }=\left(
\begin{array}{ccc}
\fr{K_3}{\sqrt{K_3^2+N_3^2+1}} & \fr{K_2}{\sqrt{K_2^2+N_2^2+1}} & \fr{%
iK_1}{\sqrt{K_1^2+N_1^2+1}} \\
\fr{N_3}{\sqrt{K_3^2+N_3^2+1}} & \fr{N_2}{\sqrt{K_2^2+N_2^2+1}} & \fr{%
iN_1}{\sqrt{K_1^2+N_1^2+1}} \\
\fr{1}{\sqrt{K_3^2+N_3^2+1}} & \fr{1}{\sqrt{K_2^2+N_2^2+1}} & \fr{i}{%
\sqrt{K_1^2+N_1^2+1}}%
\end{array}%
\right) \hspace{0.1cm}\mbox{for IH,}%
\end{array}%
\right.  \label{Unu}
\ee
where $m_{2,3}$ and $K_{1,2,3}, N_{1,2,3}$ are respectively given in Eqs. (%
\ref{m123}) and (\ref{KN123}).

The final leptonic mixing matrices then read:
  \be
U=U^+_{L} U_{\nu}=\left\{
\begin{array}{l}
\left(
\begin{array}{ccc}
\fr{K_1 \cos\theta + N_1 \sin\theta.e^{-i\al}}{\sqrt{K_1^2+N^2_1+1}} &
\fr{K_2 \cos\theta + N_2 \sin\theta. e^{-i\al}}{\sqrt{K_2^2+N^2_2+1}} &
\fr{K_3 \cos\theta + N_3 \sin\theta.e^{-i \al}}{\sqrt{K_3^2+N^2_3+1}}\\
\fr{N_1 \cos\theta - K_1 \sin\theta.e^{i\al}}{\sqrt{K_1^2+N^2_1+1}} &
\fr{N_2\cos\theta-K_2\sin\theta.e^{i \al}}{\sqrt{K^2_2+N^2_2+1}} &
\fr{N_3 \cos\theta - K_3 \sin\theta.e^{i\al}}{\sqrt{K^2_3+N^2_3+1}} \\
\fr{1}{\sqrt{K_1^2+N^2_1+1}} & \fr{1}{\sqrt{K_2^2+N^2_2+1}} & \fr{1}{%
\sqrt{K_3^2+N^2_3+1}} \\
\end{array}%
\right).P \,\,\,\,\mbox{for \, NH}, \\
\left(
\begin{array}{ccc}
\fr{K_3 \cos\theta + N_3 \sin\theta.e^{-i\al}}{\sqrt{K_3^2+N^3_2+1}} &
\fr{K_2 \cos\theta + N_2 \sin\theta.e^{-i\al}}{\sqrt{K_2^2+N^2_2+1}} &
\fr{K_1 \cos\theta + N_1 \sin\theta.e^{-i\al}}{\sqrt{K_1^2+N^2_1+1}} \\
\fr{N_3\cos\theta-K_3 \sin\theta.e^{i \al}}{\sqrt{K^2_3+N^2_3+1}} &
\fr{N_3 \cos\theta - K_2 \sin\theta.e^{i\al}}{\sqrt{K^2_2+N^2_2+1}} &
\fr{N_1 \cos\theta - K_1 \sin\theta.e^{i\al}}{\sqrt{K_1^2+N^2_1+1}} \\
\fr{1}{\sqrt{K_3^2+N^3_2+1}} & \fr{1}{\sqrt{K_2^2+N^2_2+1}} & \fr{1}{%
\sqrt{K_1^2+N^2_1+1}} \\
\end{array}%
\right).P \hspace{0.15cm}\mbox{for \ IH}.%
\end{array}%
\right.  \label{ULepg}
\ee
In the three neutrino oscillation picture\cite{Zyla:2020zbs}, the leptonic
mixing angles $\theta_{12}, \theta_{23}, \theta_{13}$ can be defined via the
elements of the leptonic mixing matrix as\footnote{%
Here, $c_{ij}=\cos \theta_{ij}$, $s_{ij}=\sin \theta_{ij}$ with $\theta_{12}$%
, $\theta_{23}$ and $\theta_{13}$ being the solar, atmospheric and reactor
angle, respectively.}:
\bea
s^2_{13}&=& |U_{13}|^2= \left\{
\begin{array}{l}
\fr{K_3^2 \cos^2\theta + K_3 N_3 \cos(2\theta)\cos\al + N_3^2
\sin^2\theta}{1 + K^2_3 + N^2_3} \hspace{2.1cm}\mbox{for \, NH}, \\
\fr{K_1^2\cos^2\theta + K_1 N_1 \sin(2\theta)\cos\al + N_1^2
\sin^2\theta}{1 + K_1^2 + N_1^2} \hspace{2.1cm}\mbox{for \ IH},%
\end{array}%
\right.  \crn
t^2_{12} &=& \left|\fr{U_{12}}{U_{11}}\right|^2=\left\{
\begin{array}{l}
\fr{(1 + K_1^2 + N_1^2) (K_2^2 \cos^2\theta + K_2 N_2
\sin(2\theta)\sin\al + N_2^2 \sin^2\theta)}{(1 + K_2^2 + N_2^2) (K_1^2
\cos^2\theta + K_1 N_1 \sin(2\theta)\sin\al + N_1^2 \sin^2\theta)}
\,\,\,\,\mbox{for \, NH}, \\
\fr{(1 + K_3^2 + N_3^2) (K_2^2 \cos^2\theta + K_2 N_2
\sin(2\theta)\cos\al + N_2^2 \sin^2\theta)}{(1 + K_2^2 + N_2^2) (K_3^2
\cos^2\theta + K_3 N_3 \sin(2\theta)\cos\al + N_3^2 \sin^2\theta)} \,\,
\hspace{0.15cm}\mbox{for \, IH},%
\end{array}%
\right.  \crn
t^2_{23}&=&\left|\fr{U_{23}}{U_{33}}\right|^2=\left\{
\begin{array}{l}
N_3^2 \cos^2\theta-K_3 N_3 \sin(2\theta) \cos\al + K_3^2 \sin^2\theta
\hspace{0.6cm} \mbox{for \, NH}, \\
N_1^2 \cos^2\theta-K_1 N_1 \sin(2\theta) \cos\al + K_1^2 \sin^2\theta
\,\, \hspace{0.5cm}\mbox{for \, IH}.%
\end{array}%
\right.  \label{thetaijsq}
\eea
In fact, the neutrino mass spectrum is currently unknown and it can have a normal or
inverted ordering depending on the sign of $\De  m^2_{31}$ (or $\De
m^2_{32}$)\cite{Zyla:2020zbs}. As will see below, the lepton
mixing matrices in Eq.(\ref{ULep}) and neutrino masses in Eq.(\ref{m123})
can fit the observed neutrino mass and mixing pattern taken from Ref.\cite%
{Esteban:2020cvm} for both Normal and Inverted hierarchies.

The Jarlskog invariant $J_{CP}$  associated with the Dirac
phase $\de _{CP}$, which controls the magnitude of CP violation effects in
neutrino oscillations \cite{Zyla:2020zbs}, is given by: 
\bea
J_{CP}&=&\mathrm{Im} (U_{23} U^*_{13} U_{12} U^*_{22})=\left\{
\begin{array}{l}
\fr{(K_2 N_3-K_3 N_2)(K_2 K_3+ N_2 N_3)}{(1 + K_2^2 + N_2^2) (1 + K_3^2 +
N_3^2)}\sin\theta\cos\theta \sin\al \hspace{0.6cm} \mbox{for \, NH}, \\
\fr{(K_2 N_1-K_1 N_2)(K_1 K_2+ N_1 N_2)}{(1 + K_1^2 + N_1^2) (1 + K_2^2 +
N_2^2)}\sin\theta\cos\theta \sin\al \hspace{0.6cm}\mbox{for \, IH}.%
\end{array}%
\right.  \label{JN}
\eea
where Eqs. (\ref{relation}) and (\ref{ULepg}) were taken into account. 

\begin{figure}[ht]
\begin{center}
\includegraphics[width=0.475\textwidth]{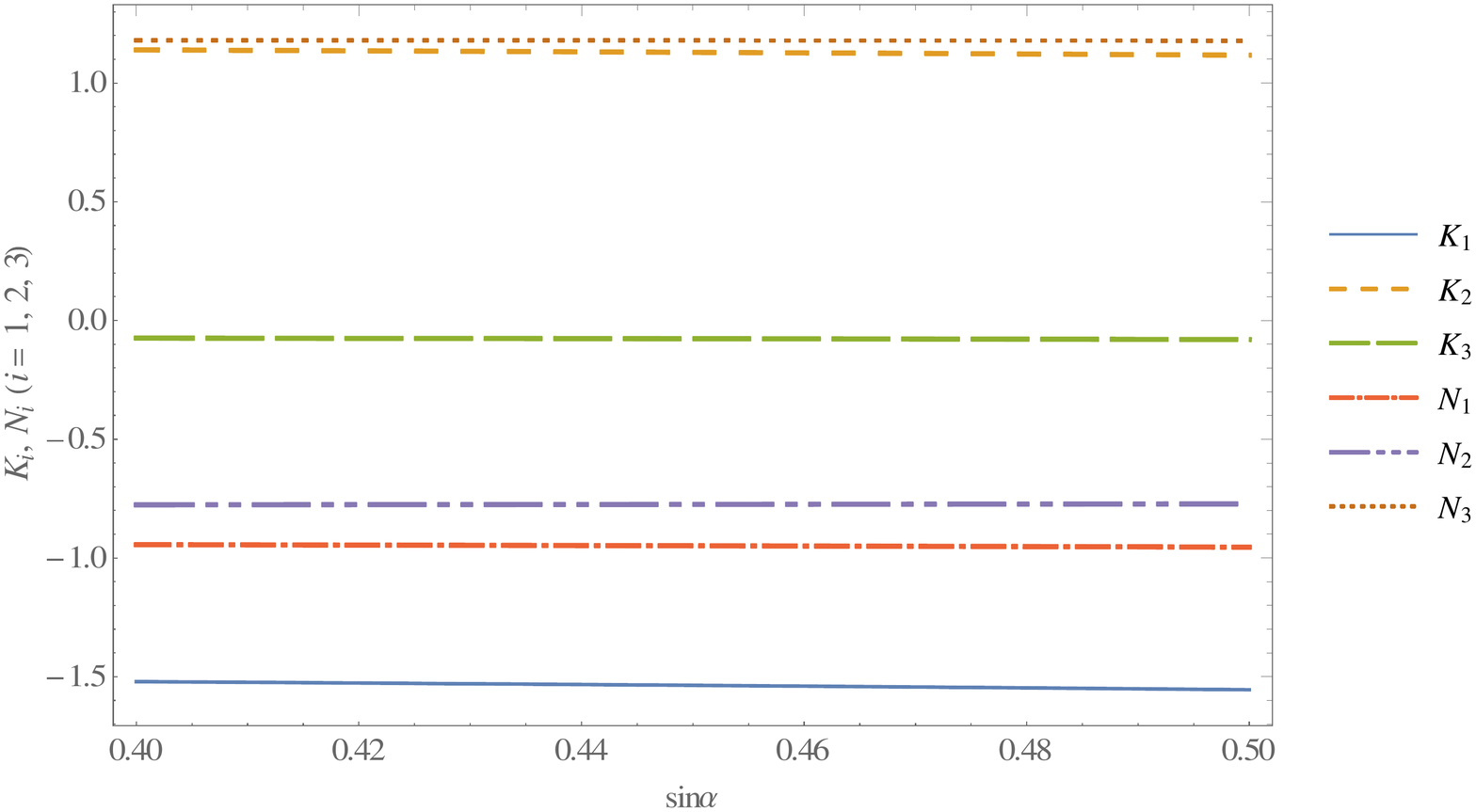} \hspace{-0.15 cm}
\includegraphics[width=0.475\textwidth]{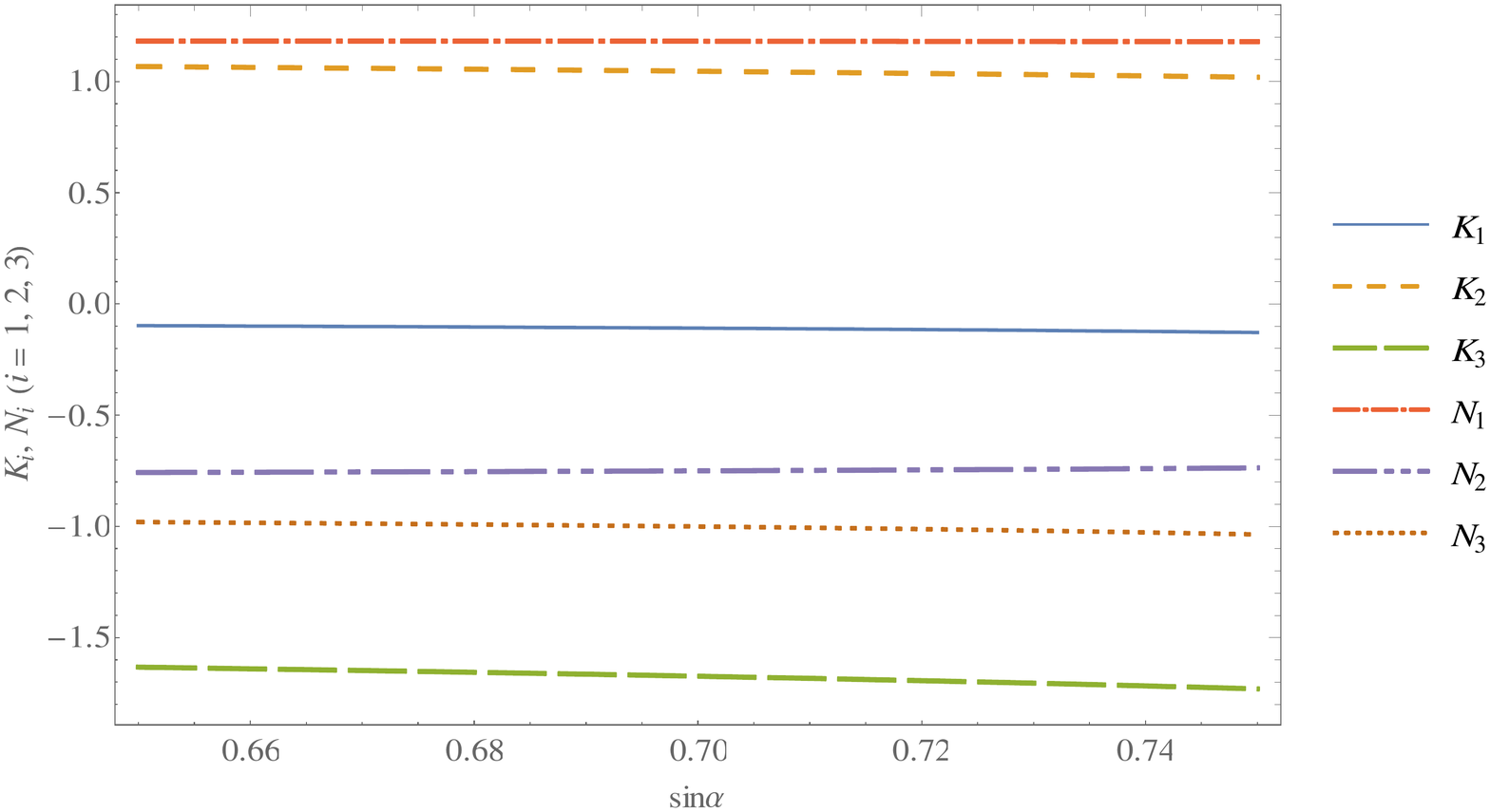}
\end{center}
\par
\vspace{-1.5 cm} \vspace{0.5 cm}
\caption{$K_{i},\, N_{i}$ versus $\sin\protect\al$ with $\sin\protect%
\al \in (0.40,\, 0.50)$ for NH (left figure) and $\sin\protect\al \in
(0.65,\, 0.75)$ for IH (right figure)}
\label{KNij}
\end{figure}

By comparing Eq. (\ref{JN}) with its corresponding expression taken from Ref.%
\cite{Zyla:2020zbs}, $J_{CP}=s_{13} c_{13}^2 s_{12} c_{12} s_{23} c_{23}
\sin\de $, we get:
\bea
\sin\de &=&\left\{
\begin{array}{l}
\fr{(K_2 N_3-K_3 N_2)(K_2 K_3+ N_2 N_3)}{(1 + K_2^2 + N_2^2) (1 + K_3^2 +
N_3^2)}\fr{\sin \theta \cos \theta \sin\al }{s_{13} c_{13}^2 s_{12}
c_{12} s_{23} c_{23}} \hspace{0.6cm} \mbox{for \, NH}, \\
\fr{(K_2 N_1-K_1 N_2)(K_1 K_2+ N_1 N_2)}{(1 + K_1^2 + N_1^2) (1 + K_2^2 +
N_2^2)}\fr{\sin \theta \cos \theta \sin\al }{s_{13} c_{13}^2 s_{12}
c_{12} s_{23} c_{23}} \hspace{0.6cm}\mbox{for \, IH}.%
\end{array}%
\right.  \label{sdN}
\eea %
\newline
Combining Eqs. (\ref{relation}), (\ref{ULepg}) and (\ref{thetaijsq}), we
found that $N_i, K_i \, (i=1,2,3)$ and all the elements of leptonic mixing
matrix U depend on five parameters $\theta_{12}, \theta_{13}, \theta_{23}$, $%
\theta$ and $\al $. By using the best-fit values of leptonic mixing
angles taken from Ref. \cite{Esteban:2020cvm},
\bea
&&\sin^2 \theta_{12}=0.304,\hspace*{0.2cm} \sin^2 \theta_{23}=\left\{
\begin{array}{l}
0.573 \hspace{0.3cm} \mbox{for \, NH}, \\
0.575 \hspace{0.3cm}\mbox{for \, IH},%
\end{array}%
\right. \hspace*{0.2cm} \sin^2 \theta_{13}=\left\{
\begin{array}{l}
0.02219 \hspace{0.3cm} \mbox{for \, NH}, \\
0.02238 \hspace{0.3cm}\mbox{for \, IH},%
\end{array}%
\right.
\eea
we find out the regions of $\theta$ and $\al $ 
which can reproduce the recent experimental data. Indeed, in the case where $%
\sin\theta=0.25\, (\theta=14.5^\circ)$, $N_i$ and $K_i$ depend on
$\al $ with $\sin\al \in (0.40, 0.50)$ for NH and $\sin\al \in
(0.65, 0.75)$ for IH which is plotted in Fig. \ref{KNij}.

Next, taking the best-fit values of the solar and atmospheric neutrino
squared-mass differences taken form Ref.\cite{Esteban:2020cvm},
\bea
&&\De  m^2_{21}=7.42 \times 10^{-5} \,\mathrm{eV}^2, \hspace*{0.2cm}
\left\{
\begin{array}{l}
\De  m^2_{31}=2.517 \times 10^{-3}\, \mathrm{eV}^2 \hspace{0.375cm} %
\mbox{for  NH}, \\
\De  m^2_{32}=-2.498 \times 10^{-3}\, \mathrm{eV}^2 \hspace{0.1cm}%
\mbox{for  IH},%
\end{array}%
\right.
\eea
we get a solution
\bea
&&\Bbbk_1=\left\{
\begin{array}{l}
2.94\times 10^{-2}\, \mathrm{eV} \hspace{0.2cm} \mbox{for  NH}, \\
4.96 \times 10^{-2}\, \mathrm{eV} \hspace{0.2cm}\mbox{for  IH},%
\end{array}%
\right. \hspace*{0.2cm} \Bbbk_2=\left\{
\begin{array}{l}
2.08\times 10^{-2}\, \mathrm{eV} \hspace{0.35cm} \mbox{for  NH}, \\
-3.74 \times 10^{-4}\, \mathrm{eV} \hspace{0.1cm}\mbox{for  IH},%
\end{array}%
\right.
\eea
and three neutrino masses are explicitly given as
\bea
m_1&=& 0 \, \mathrm{eV},\, \, m_2= 8.61\times 10^{-3} \, \mathrm{eV}, \, \,
m_3= 5.02\times 10^{-2}\, \mathrm{eV} \hspace*{0.2cm} \mathrm{for \hspace*{%
0.2cm} NH},  \label{m1m2m3N} \\
m_1&=& 4.92 \times 10^{-2}\, \mathrm{eV},\, \, m_2= 5.0 \times 10^{-2}\,
\mathrm{eV}, \, \, m_3= 0 \, \mathrm{eV} \hspace*{0.2cm} \mathrm{for
\hspace*{0.2cm} IH}.  \label{m1m2m3I}
\eea
The sum of three neutrino masses is thus found to be
\bea
\sum_{i=1}^3 m_i=\left\{
\begin{array}{l}
5.88\times 10^{-2} \,\mathrm{eV} \hspace{0.2cm} \mbox{for NH}, \\
9.92\times 10^{-2} \,\mathrm{eV} \hspace{0.2cm}\mbox{for IH},%
\end{array}%
\right.
\eea
which are well consistent with the updated bounds from cosmology \cite%
{RoyChoudhury:2019hls}.

Furthermore, in the NH, $m_1\approx m_2<m_3$, so $m_{1}=0$ is the lightest
neutrino mass while $m_{3}=0$ is the lightest neutrino mass for IH. The
effective neutrino mass parameters governing the beta decay and
neutrinoless double beta decay, $m_{\beta }= \sqrt{\sum^3_{i=1}
\left|U_{1i}\right|^2 m_i^2}$ and $\langle m_{ee}\rangle=\left| \sum^3_{i=1}
U_{1i}^2 m_i \right|$ depend only on $\sin\al$ with $\sin\al \in
(0.40,\, 0.50)$ for NH and 
$\sin\al \in (0.65,\, 0.75)$ for IH which is depicted in Fig.\ref%
{membD4e331}.
\begin{figure}[ht]
\begin{center}
\vspace{0.25 cm} 
\includegraphics[width=0.45\textwidth]{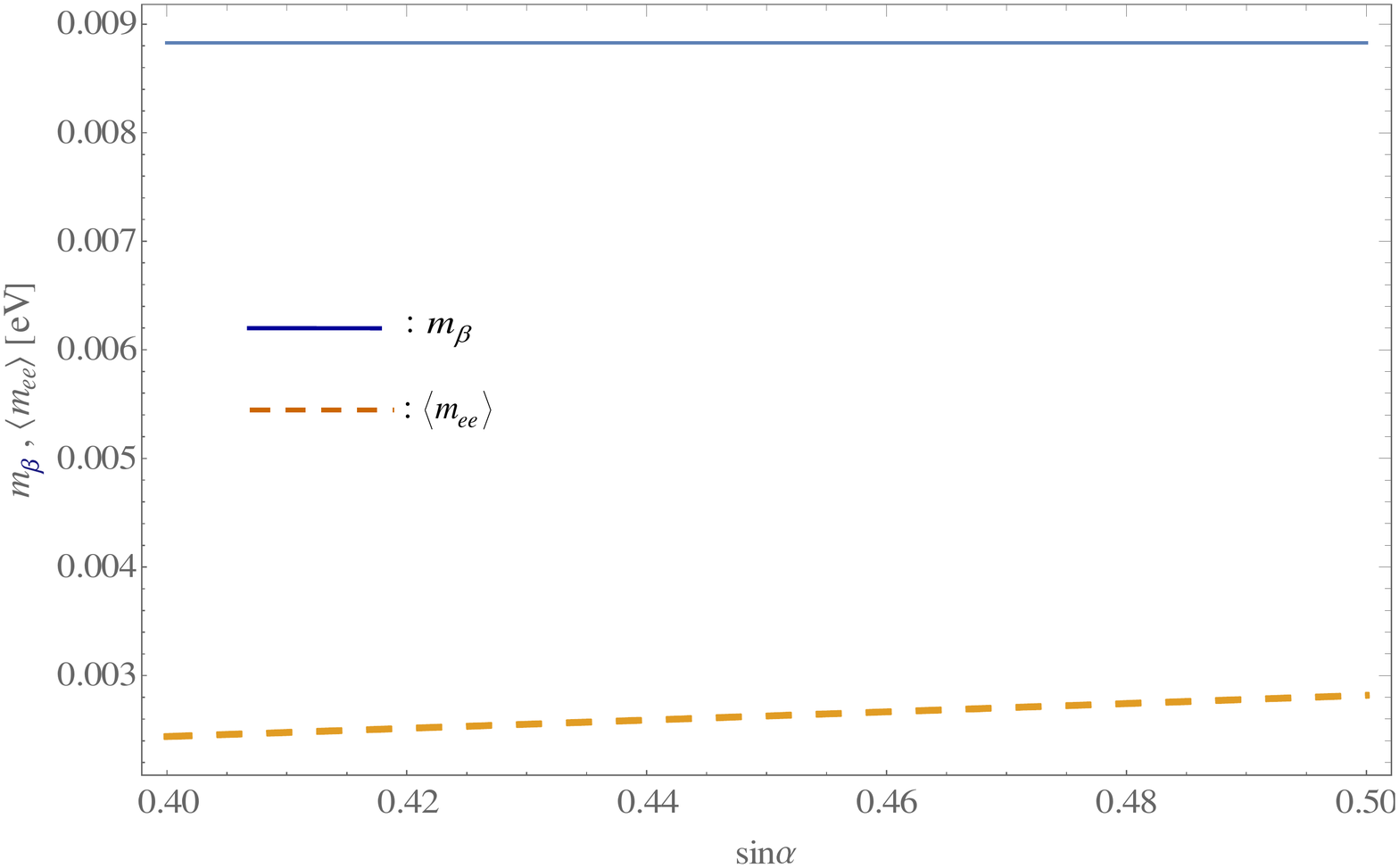} \hspace{-0.2 cm} %
\includegraphics[width=0.45\textwidth]{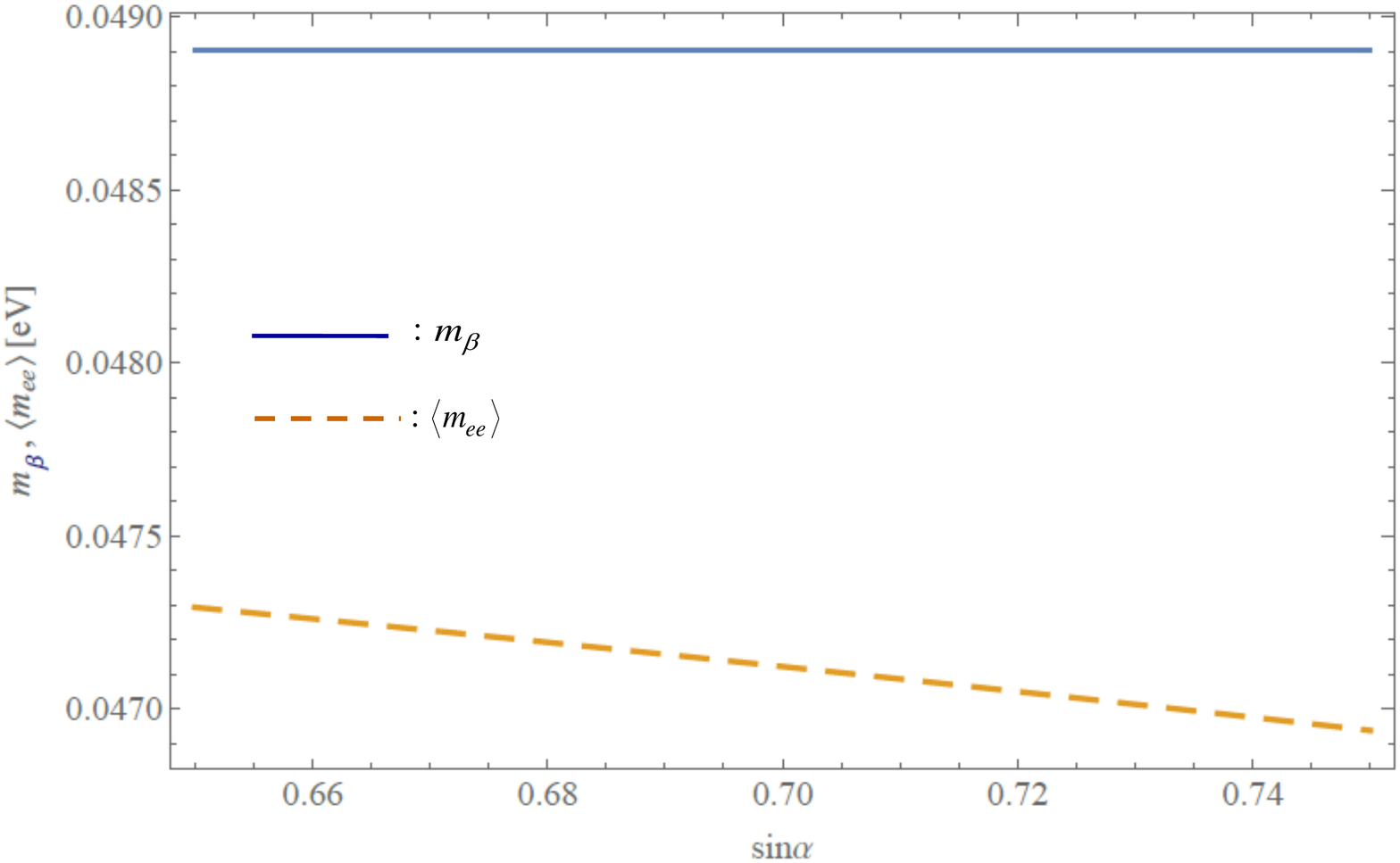}
\end{center}
\par
\vspace{-1.0 cm}
\caption{$\langle m_{ee}\rangle$ and $m_{\protect\beta }$ versus $\sin%
\protect\al$ with $\sin\protect\al \in (0.40,\, 0.50)$ for NH (left
figure) and $\sin\protect\al \in (0.65,\, 0.75)$ for IH (right figure)}
\label{membD4e331}
\end{figure}
In the case where $\sin\al =0.445\, (\al=26.4^\circ)$ for NH and $\sin\al
=0.75\, (\al=48.6^\circ)$ for IH, $m_{\beta }$ and $\langle m_{ee}\rangle$
are found to be:
\bea
\langle m_{ee}\rangle &=&\left\{
\begin{array}{l}
2.61 \times 10^{-3} \,\mathrm{eV} \hspace{0.3cm} \mathrm{for}\hspace{0.1cm}
\mathrm{NH}, \\
4.69\times 10^{-2} \,\mathrm{eV} \hspace{0.3cm} \mathrm{for} \hspace{0.1cm}
\mathrm{IH},%
\end{array}%
\right.  \label{mbetvalues}
\eea
and
\bea
m_{\beta} =\left\{
\begin{array}{l}
8.83 \times 10^{-3} \,\mathrm{eV} \hspace{0.3cm} \mathrm{for}\hspace{0.1cm}
\mathrm{NH}, \\
4.89\times 10^{-2} \,\mathrm{eV} \hspace{0.3cm} \mathrm{for} \hspace{0.1cm}
\mathrm{IH}.%
\end{array}%
\right.
\eea
The Jarlskog invariant $J_{CP}$, determined from Eq. (\ref{JN}%
), possessed the following values:
\bea
J_{CP}=\left\{
\begin{array}{l}
-0.0199 \hspace{0.3cm} \mathrm{for}\hspace{0.1cm} \mathrm{NH}, \\
-0.0323 \hspace{0.3cm} \mathrm{for} \hspace{0.1cm} \mathrm{IH}.%
\end{array}%
\right.
\eea
The lepton mixing matrices for both Normal and Inverted hierarchies take the
explicit forms
  \be
U=\left\{
\begin{array}{l}
\left(
\begin{array}{ccc}
-0.823+0.0511i & 0.543+0.0508i & 0.0847+0.123 i \\
-0.279+0.0828 i & -0.591-0.0741i & -0.0051+0.749 i \\
0.481 & 0.589 & 0.646i%
\end{array}%
\right) \hspace{0.2 cm}\mbox{for \, NH}, \\
\left(
\begin{array}{ccc}
-0.82+0.0863i & \hspace*{0.2cm} 0.538+0.086i & \hspace*{0.2cm} 0.143+0.0455i
\\
-0.319+0.144 i & \hspace{0.25cm}-0.549-0.119i & \hspace{0.25cm} -0.0155+0.75i
\\
0.444 & 0.622 & 0.645i%
\end{array}%
\right) \hspace{0.25cm}\mbox{for \, IH},%
\end{array}%
\right.  \label{ULep}
\ee
which are unitary and consistent with the constraint on the absolute values
of the entries of the lepton mixing matrix given in Ref.\cite%
{Esteban:2020cvm}. The other model parameters are obtained as in Tab.\ref%
{paraD4}.
\begin{table}[ht]
\caption{The model parameters in the case $\sin\protect\theta=0.25$, and $%
\sin\protect\al =0.445$ for NH\, and $\sin\protect\theta=0.25$ and $\sin%
\protect\al =0.75$ for IH}
\label{paraD4}{%
\begin{tabular}{@{}cccccccccc}
\hline
& Parameters & \hspace*{0.2cm} The derived values (NH) & \hspace*{0.2cm} The
derived values (IH) &  &  &  &  &  &  \\ \hline
& \,\,$K_1$ & \, $-1.53 $ & \, $-0.128 $ &  &  &  &  &  &  \\
& \,\,$K_2$ & \, $1.13$ & \, $1.02$ &  &  &  &  &  &  \\
& \,\,$K_3$ & \, $-0.0767 $ & \, $-1.73 $ &  &  &  &  &  &  \\
& \,\,$N_1$ & \, $-0.948$ & \, $1.18$ &  &  &  &  &  &  \\
& \,\, $N_2$ & \, $-0.775$ & \, $-0.737$ &  &  &  &  &  &  \\
& \,\, $N_3$ & \, $1.18$ & \, $-1.04$ &  &  &  &  &  &  \\ \hline
\end{tabular}%
}
\end{table}

\section{Scalar potential with two $SU\left(3\right)_{L}$ triplets}

\label{scalars} To simplify our analysis, we neglect the mixing terms
between the $SU\left(3\right)_{L}$ scalar triplets and the gauge singlet
scalars. Then, the scalar potential for the two $SU\left(3\right)_{L}$
scalar triplets is given by:
\bea
V &=& -\mu_{\chi }^{2}(\chi^{\dagger}\chi)
-\mu_{\rho}^{2}(\rho^{\dagger}\rho) + \la
_{1}(\chi^{\dagger}\chi)(\chi^{\dagger}\chi) + \la
_{2}(\rho^{\dagger}\rho)(\rho^{\dagger}\rho) + \la
_{3}(\chi^{\dagger}\chi)(\rho^{\dagger}\rho) + \la
_{4}(\chi^{\dagger}\rho)(\rho^{\dagger}\chi) + H.c.,\nn
\eea
with $\chi$ and $\rho$, the $SU(3)_{L}$ scalar triplets. Furthermore, the
global minimum conditions of the scalar potential give the 
relations:
\bea
\mu_{\chi}^{2} &=& \dfrac{1}{2} \left(2v_{\chi}^{2} \la  _{1} +
v_{\rho}^{2} \la  _{3} \right), \\
\mu_{\rho}^{2} &=& \dfrac{1}{2} \left(2v_{\rho}^{2} \la  _{2} +
v_{\chi}^{2} \la  _{3} \right).
\eea

After spontaneous symmetry breaking we get the squared mass matrices for the
scalar fields:
\bea
M_{CPeven}^{2} &=& \left(
\begin{array}{ccc}
0 & 0 & 0 \\
0 & v_{\rho}^{2} \la  _{2} & \dfrac{1}{2} v_{\rho} v_{\chi} \la  _{3}
\\
0 & \dfrac{1}{2} v_{\rho} v_{\chi} \la  _{3} & v_{\chi}^{2} \la  _{1}%
\end{array}
\right), \hspace{0.05cm} M_{CPodd}^{2} = \left(
\begin{array}{ccc}
0 & 0 & 0 \\
0 & 0 & 0 \\
0 & 0 & 0%
\end{array}
\right), \hspace{0.05cm}M_{charged}^{2} = \left(
\begin{array}{ccc}
0 & 0 & 0 \\
0 & \dfrac{1}{2} v_{\rho}^{2} \la  _{4} & \dfrac{1}{2} v_{\rho} v_{\chi}
\la  _{4} \\
0 & \dfrac{1}{2} v_{\rho} v_{\chi} \la  _{4} & \dfrac{1}{2} v_{\chi}^{2}
\la  _{4}%
\end{array}
\right).
\eea
This shows that the resulting physical scalar spectrum arising from the $%
SU(3)_L$ scalar triplets $\chi$ and $\rho$ is composed of the $126$ GeV SM
like Higgs boson, a heavy neutral CP even scalar $H^0$ associated with the
spontaneous breaking of the $SU(3)_L\times U(1)_X$ symmetry and the
electrically charged scalars $H^{\pm}$. The massless degrees of freedom in
the scalar spectrum correspond to the Goldstone boson associated with the
longitudinal components of the $W^{\pm}$, $Z$, $W^{\prime\pm}$, $Z^{\prime}$%
, $K^{0}$ and $\bar{K}^{0}$ massive gauge bosons.

\section{Higgs diphoton decay rate constraints}

\label{diphoton} The decay rate for the $h\rightarrow \ga  \ga  $
process takes the form {\cite{Shifman:1979eb,Gavela:1981ri,Kalyniak:1985ct,Spira:1997dg,Djouadi:2005gj,Marciano:2011gm,Wang:2012gm}:}
\bea
\Ga  (h \rightarrow \ga  \ga  ) &=&\dfrac{\al  _{em}^2 m_h^3}{256
\pi^3 v^2}\left|\sum_f a_{hff} N_C Q_f^2 F_{1/2}(\rho_f)+a_{hWW}
F_{1}(\rho_W)+\fr{C_{hH^{\pm}H^{\mp}}}{2m^{2}_{H^{\pm}}}%
vF_{0}(\rho_{H^{\pm}})\right|^2,
\eea

where $\rho_i$ are the mass ratios $\rho_i= \fr{m_h^2}{4 M_i^2}$ with $%
M_i=m_f, M_W$; $\al  _{em}$ is the fine structure constant; $N_C$ is the
color factor ($N_C=1$ for leptons and $N_C=3$ for quarks) and $Q_f$ is the
electric charge of the fermion in the loop. From the fermion-loop
contributions we only consider the dominant top quark term. Furthermore, $%
C_{hH^{\pm}H^{\mp}}$ is the trilinear coupling between the SM-like Higgs and
a pair of charged Higgs bosons, whereas $a_{htt}$ and $a_{hWW}$ are the
deviation factors from the SM Higgs top quark coupling and the SM Higgs-W
gauge boson coupling, respectively (in the SM these factors are unity). Such
deviation factors are very close to unity in our model, which is a
consequence of the numerical analysis of its scalar, Yukawa and gauge
sectors. Besides, $F_{1/2}(z)$ and $F_{1}(z)$ are the dimensionless loop
factors for spin-$1/2$ and spin-$1$ particles running in the internal lines
of the loops. They are given by:

\begin{align}
F_{1/2}(z) &= 2(z + (z -1)f(z))z^{-2}, \\
F_{1}(z) &= -2(2z^2 + 3z + 3(2z-1)f(z))z^{-2}, \\
F_{0}(z) &= -(z - f(z))z^{-2},
\end{align}
with
\begin{align}
f(z) = \left\{
\begin{array}{lcc}
\left(\arcsin \sqrt{2}\right)^2 & \text{for} & z \leq 1, \\
&  &  \\
-\fr{1}{4}\left(\ln \left(\fr{1+\sqrt{1-z^{-1}}}{1-\sqrt{1-z^{-1}}-i\pi}
\right)^2 \right) & \text{for} & z > 1.%
\end{array}
\right.
\end{align}

In order to study the implications of our model in the decay of the $126$
GeV Higgs boson into a photon pair, one introduces the Higgs diphoton signal
strength $R_{\ga  \ga  }$, which is defined as:
\begin{align}
R_{\ga  \ga  } = \fr{\si (pp \to h)\Ga  (h \to \ga  \ga  )}{%
\si (pp \to h)_{SM}\Ga  (h \to \ga  \ga  )_{SM}} \simeq a^{2}_{htt}
\fr{\Ga  (h \to \ga  \ga  )}{\Ga  (h \to \ga  \ga  )_{SM}}.
\label{eqn:hgg}
\end{align}

That Higgs diphoton signal strength, normalizes the $\ga  \ga  $ signal
predicted by our model in relation to the one given by the SM. Here we have
used the fact that in our model, single Higgs production is also dominated
by gluon fusion as in the Standard Model.

The ratio $R_{\ga  \ga  }$ has been measured by CMS and ATLAS
collaborations with the best fit signals \cite{Sirunyan:2018ouh,Aad:2019mbh}:

\begin{align}
R^{CMS}_{\ga  \ga  } = 1.18^{+0.17}_{-0.14} \quad \text{and} \quad
R^{ATLAS}_{\ga  \ga  } = 0.96 \pm 0.14.  \label{eqn:rgg}
\end{align}

\begin{figure}[]
\begin{center}
\includegraphics[scale=0.450]{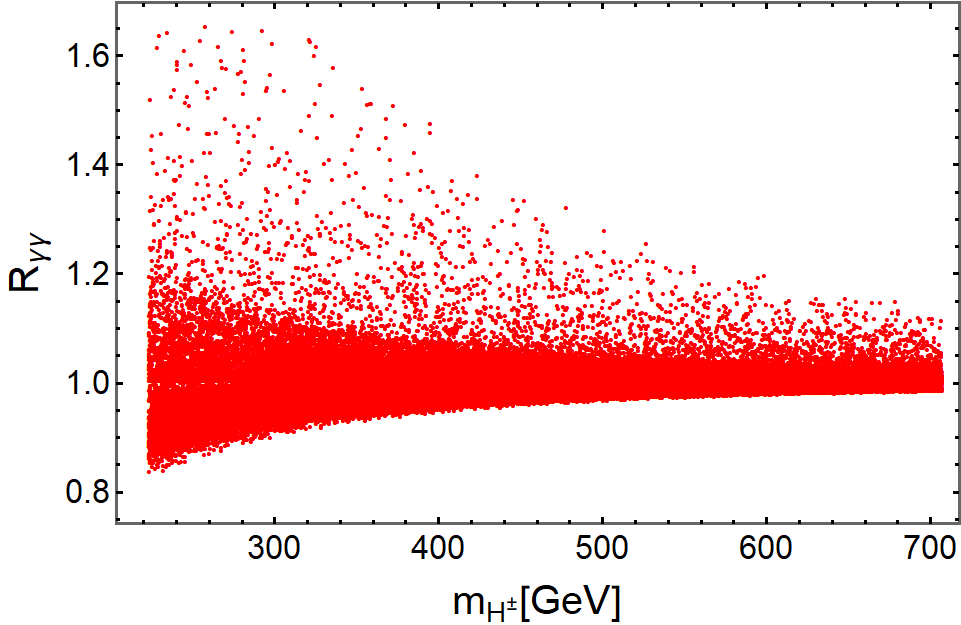}
\end{center}
\caption{Higgs diphoton signal strength as a function of the charged scalar
mass $m_{H^{\pm}}$, with $600\,[GeV] < m_{H^{\pm}} < 7000\,[GeV]$} 
\label{fig:Diphoton}
\end{figure}

The Higgs diphoton signal strength as a function of the electrically charged
scalar mass is shown in Figure \ref{fig:Diphoton}. This shows that our model
successfully accommodates the current Higgs diphoton decay rate constraints.


\section{Muon anomalous magnetic moment}
\label{gminus2}
\begin{figure}[h]
\vspace{0.5 cm}
\includegraphics[width=0.5\textwidth]{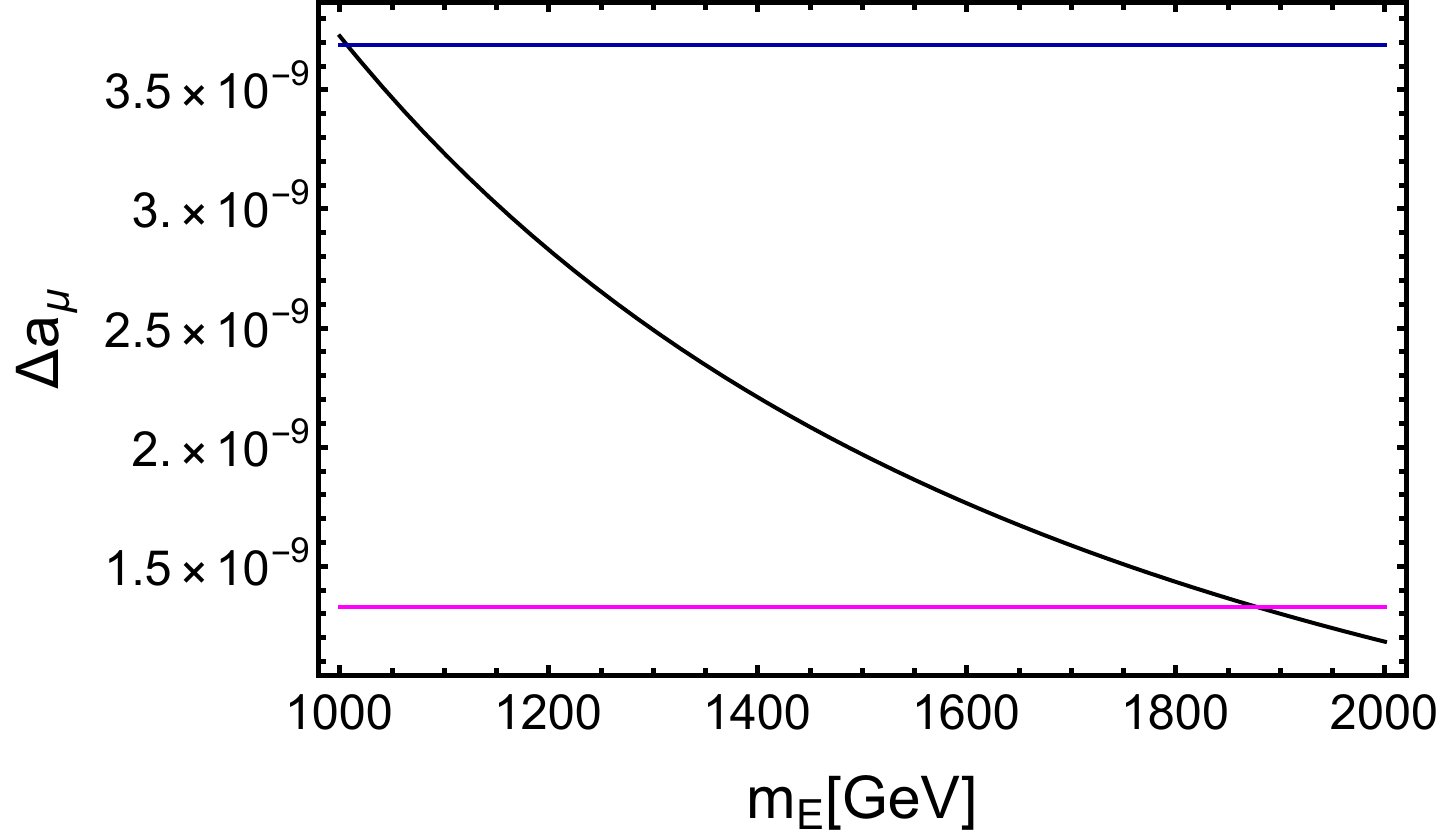}
\caption{Muon anomalous magnetic moment as a function of the charged exotic lepton mass $m_E$.}
\label{gminus2muonvsmE}
\end{figure}

  {In this section we discuss the implications of our model in the muon anomalous magnetic moment. It is worth mentioning that the dominant contribution to the muon anomalous magnetic moment arises from the one-loop diagram involving the exchange of the electrically neutral scalars $h$ and $H$ and the charged exotic vector like leptons $E_1$ and $E_2$.
\crb{It is worth mentioning that there other contributions to the muon anomalous magnetic moments like the ones arising from the virtual exchange of heavy neutral and electrically charged gauge bosons together with charged and neutral leptons, respectively as well as contributions due to electrically charged scalars and neutrinos. However those extra contributions are subleading. Regarding the contribution arising from the virtual exchange electrically charged scalars and light active neutrinos, we have numerically checked that it can reproduce the magnitude of the $g-2$ muon anomaly for electrically charged scalars lower than $400$ GeV. However such contribution turn out to be negative and thus not allow to reproduce the correct sign of the $g-2$ muon anomaly. Consequently, in our analysis of the muon anomalous magnetic moment, we only consider the leading contribution arising from the virtual exchange of electrically neutral scalars $h$ and $H$ and the charged exotic vector like leptons $E_1$ and $E_2$. Furthermore, in order to simplify our numerical analysis, we restrict to the region of parameter space where the electrically charged scalars are heavier than about $400$ GeV, thus implying that the contribution to the $g-2$ muon anomaly arising from the virtual exchange electrically charged scalars and light active neutrinos is suppressed and therefore subleading. The Feynman diagrams corresponding the Beyond Standard Model contributions to the muon anomalous magnetic moment in the 3-3-1 model under consideration are shown in Fig.\ref{k3F}.}
\begin{figure}[h]
\begin{center}
\vspace{-1.5 cm}
\hspace{-5.0 cm}\includegraphics[width=1.0\textwidth]{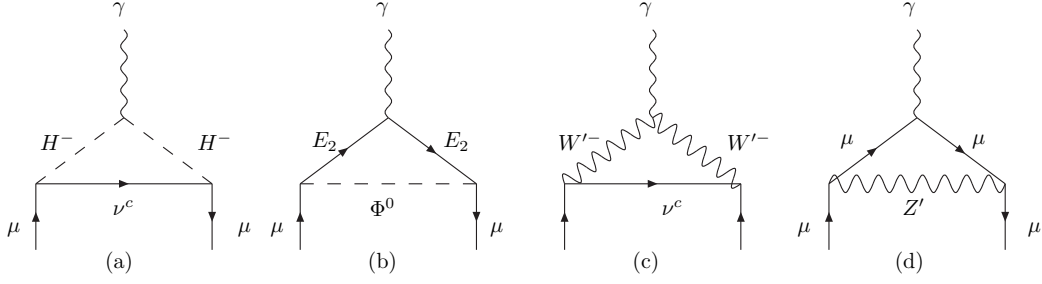}\hspace*{-3.5 cm}
\end{center}
\vspace{-17.65 cm}
\caption{\crb{Feynman diagrams corresponding to Beyond Standard Model contributions to the muon anomalous magnetic moment in the 3-3-1 model under consideration. Notice that the second diagram involving the virtual exchange of charged exotic leptons $E_{1,2}$ and neutral scalars $\Phi^0=h,H$ is the one that provides the leading contribution to the muon anomalous magnetic moment.}}
\label{k3F}
\end{figure}
\crb{In view of the previous discussion,} the dominant contribution to the muon anomalous magnetic moment in our model has the form:
  \be
\De  a_{\mu }\simeq \fr{y_{1}^{\left( l\right) }z_{2}^{\left( l\right)
}m_{\mu }^{2}}{8\pi ^{2}}\left[ J\left( m_{E_{2}},m_{h}\right) -J\left(
m_{E_{2}},m_{H}\right) \right] \sin \theta \cos \theta ,
\ee %
where, $y_{1}^{\left( l\right) }$ and $z_{2}^{\left( l\right)}$ are the leptonic Yukawa couplings appearing in the first line of Eq. (\ref{Lyl}). Here, in order to simplify our analysis we have restricted to the case $v_1 \ll v_2$, which implies that only $\phi_1$ (the first component of the $D_4$ scalar doublet $\Phi$) mixes with the CP even neutral part of the $SU(3)_L$ scalar triplet $\rho$. Then, the neutral scalars $h$ and $H$ are defined as:  $H\simeq \cos \theta \func{Re}\phi _{1}+\sin \theta \xi _{\rho }$ , $%
h\simeq -\sin \theta \func{Re}\phi _{1}+\cos \theta \xi _{\rho }$, and $%
m_{E_{2}}$ is the mass of the VLL $E_{2}$. Furthermore, the loop $J\left(
m_{E},m_{S}\right) $ function has the following form \cite%
{Diaz:2002uk,Jegerlehner:2009ry,Kelso:2014qka,Lindner:2016bgg}:
  \be
J\left( m_{E},m_{S}\right) =\int_{0}^{1}dx\fr{x^{2}\left( 1-x+\fr{m_{E}}{%
m_{\mu }}\right) }{m_{\mu }^{2}x^{2}+\left( m_{E}^{2}-m_{\mu }^{2}\right)
x+m_{S}^{2}\left( 1-x\right) }.
\ee
The above given expression for the muon anomalous magnetic moment can be approximatelly rewritten as follows:
  \be
\De  a_{\mu }\simeq \fr{y_{1}^{\left( l\right) }z_{2}^{\left( l\right) }%
}{8\pi ^{2}}\left[ \fr{m_{\mu }m_{E_{2}}}{m_{h}^{2}}G_{1}\left( \fr{%
m_{E_{2}}^{2}}{m_{h}^{2}}\right) -\fr{m_{\mu }m_{E_{2}}}{m_{H}^{2}}%
G\left( \fr{m_{E_{2}}^{2}}{m_{H}^{2}}\right) \right] \sin \theta \cos
\theta ,  \label{amuNS}
\ee %
where the corresponding loop function as the form \cite{Kowalska:2017iqv}:
  \be
G\left(z\right) =\fr{3-4z+z^{2}+2\ln z}{2\left( z-1\right) ^{3}}.
\label{G1}
\ee

Considering that the muon anomalous magnetic moment is constrained to be in
the range \cite%
{Hagiwara:2011af,Davier:2017zfy,Blum:2018mom,Keshavarzi:2018mgv,Nomura:2018lsx,Nomura:2018vfz,Aoyama:2020ynm,Abi:2021gix}
  \be
\left( \De  a_{\mu }\right) _{\exp }=\crb{a_{\mu }^{\mathrm{exp}}-a_{\mu }^{\mathrm{SM}}}=\left( 2.51\pm 0.59\right) \times
10^{-9}.
\ee
We display in Fig.\ref{gminus2muonvsmE} the muon anomalous magnetic moment as a function of the charged exotic vector like mass.
%
As shown in Fig.\ref{gminus2muonvsmE}, we have that our model successfully accommodates the experimental value of $\De  a_{\mu }$ \crb{for charged exotic lepton masses at the TeV scale.}}

\section{Meson oscillations}  \label{FCNC}The non universal $U(1)_X$ charge
assignments for the left handed quark fields yield tree level $Z^{\prime}$
mediated flavour changing neutral processes (FCNC) which will yield $K^0-%
\bar{K}^0$, $B^0_d-\bar{B}^0_d$ and $B^0_s-\bar{B}^0_s$ meson oscillations.
These meson mixings are described by the following effective Hamiltonian
interactions \cite{Queiroz:2016gif}:
  \be
\mathcal{H}_{eff}^{\left( K^{0}-\bar{K}^{0}\right) }\mathcal{=}\fr{4\sqrt{%
2 }G_{F}c_{W}^{4}m_{Z}^{2}}{\left( 3-4s_{W}^{2}\right) m_{Z^{\prime }}^{2}}
\left\vert \left( V_{DL}^{\ast }\right) _{32}\left( V_{DL}\right)
_{31}\right\vert ^{2}O^{\left( K^{0}-\bar{K}^{0}\right) },
\ee
  \be
\mathcal{H}_{eff}^{\left( B_{d}^{0}-\bar{B}_{d}^{0}\right) }\mathcal{=}\fr{
4\sqrt{2}G_{F}c_{W}^{4}m_{Z}^{2}}{\left( 3-4s_{W}^{2}\right) m_{Z^{\prime
}}^{2}}\left\vert \left( V_{DL}^{\ast }\right) _{31}\left( V_{DL}\right)
_{33}\right\vert ^{2}O^{\left( B_{d}^{0}-\bar{B}_{d}^{0}\right) },
\ee
  \be
\mathcal{H}_{eff}^{\left( B_{s}^{0}-\bar{B}_{s}^{0}\right) }\mathcal{=}\fr{
4\sqrt{2}G_{F}c_{W}^{4}m_{Z}^{2}}{\left( 3-4s_{W}^{2}\right) m_{Z^{\prime
}}^{2}}\left\vert \left( V_{DL}^{\ast }\right) _{32}\left( V_{DL}\right)
_{33}\right\vert ^{2}O^{\left( B_{s}^{0}-\bar{B}_{s}^{0}\right) }.
\ee
where the corresponding operators are given by:
\bea
O^{\left( K^{0}-\bar{K}^{0}\right) } &=&\left( \overline{s}\ga  _{\mu
}P_{L}d\right) \left( \overline{s}\ga  ^{\mu }P_{L}d\right) ,\hspace{0.7cm}
\hspace{0.7cm}O^{\left( B_{d}^{0}-\bar{B}_{d}^{0}\right) }=\left( \overline{%
d }\ga  _{\mu }P_{L}b\right) \left( \overline{d}\ga  ^{\mu
}P_{L}b\right) , \\
O^{\left( B_{s}^{0}-\bar{B}_{s}^{0}\right) } &=&\left( \overline{s}\ga
_{\mu }P_{L}b\right) \left( \overline{s}\ga  ^{\mu }P_{L}b\right).
\eea
Furthermore, the following relations have been taken into account:
\bea
\widetilde{M}_{f} &=&\left( M_{f}\right) _{diag}=V_{fL}^{\dagger
}M_{f}V_{fR},\hspace{1cm}\hspace{1cm}f_{\left( L,R\right) }=V_{f\left(
L,R\right) }\widetilde{f}_{\left( L,R\right) },  \crn
\overline{f}_{iL}\left( M_{f}\right) _{ij}f_{jR} &=&\overline{\widetilde{f}}
_{kL}\left( V_{fL}^{\dagger }\right) _{ki}\left( M_{f}\right) _{ij}\left(
V_{fR}\right) _{jl}\widetilde{f}_{lR}=\overline{\widetilde{f}}_{kL}\left(
V_{fL}^{\dagger }M_{f}V_{fR}\right) _{kl}\widetilde{f}_{lR}=\overline{
\widetilde{f}}_{kL}\left( \widetilde{M}_{f}\right) _{kl}\widetilde{f}
_{lR}=m_{fk}\overline{\widetilde{f}}_{kL}\widetilde{f}_{kR},  \crn
k &=&1,2,3\,.
\eea
Here, $\widetilde{f}_{k\left( L,R\right) }$ and $f_{k\left( L,R\right) }$ ($
k=1,2,3$) are the SM fermionic fields in the mass and interaction bases,
respectively.

On the other hand, the $K-\bar{K}$, $B_{d}^{0}-\bar{B}_{d}^{0}$ and $
B_{s}^{0}-\bar{B}_{s}^{0}$\ mass splittings are given by:
  \be
\De  m_{K}=\left( \De  m_{K}\right) _{SM}+\De  m_{K}^{\left( NP\right)
},\hspace{1cm}\De  m_{B_{d}}=\left( \De  m_{B_{d}}\right) _{SM}+\De
m_{B_{d}}^{\left( NP\right) },\hspace{1cm}\De  m_{B_{s}}=\left( \De
m_{B_{s}}\right) _{SM}+\De  m_{B_{s}}^{\left( NP\right) },  \label{Deltam}
\ee
where $\left( \De  m_{K}\right) _{SM}$, $\left( \De  m_{B_{d}}\right)
_{SM}$ and $\left( \De  m_{B_{s}}\right) _{SM}$ are the SM contributions,
whereas $\De  m_{K}^{\left( NP\right) }$ , $\De  m_{B_{d}}^{\left(
NP\right) }$ and $\left( \De  m_{B_{s}}\right) _{SM}$ are new physics
contributions.

The new physics contributions to meson \crb{mass} differences are \cite{Queiroz:2016gif}:

  \be
\De  m_{K}^{\left( NP\right) }=\fr{4\sqrt{2}G_{F}c_{W}^{4}m_{Z}^{2}}{
\left( 3-4s_{W}^{2}\right) m_{Z^{\prime }}^{2}}\left\vert \left(
V_{DL}^{\ast }\right) _{32}\left( V_{DL}\right) _{31}\right\vert
^{2}f_{K}^{2}B_{K}\eta _{K}m_{K},
\ee
  \be
\De  m_{B_{d}}^{\left( NP\right) }=\fr{4\sqrt{2}G_{F}c_{W}^{4}m_{Z}^{2}}{
\left( 3-4s_{W}^{2}\right) m_{Z^{\prime }}^{2}}\left\vert \left(
V_{DL}^{\ast }\right) _{31}\left( V_{DL}\right) _{33}\right\vert
^{2}f_{B_{d}}^{2}B_{B_{d}}\eta _{B_{d}}m_{B_{d}},
\ee
  \be
\De  m_{B_{s}}^{\left( NP\right) }=\fr{4\sqrt{2}G_{F}c_{W}^{4}m_{Z}^{2}}{
\left( 3-4s_{W}^{2}\right) m_{Z^{\prime }}^{2}}\left\vert \left(
V_{DL}^{\ast }\right) _{32}\left( V_{DL}\right) _{33}\right\vert
^{2}f_{B_{s}}^{2}B_{B_{s}}\eta _{B_{s}}m_{B_{s}}.
\ee
Using the following parameters \cite%
{Dedes:2002er,Aranda:2012bv,Khalil:2013ixa,Queiroz:2016gif,Buras:2016dxz,Ferreira:2017tvy,Duy:2020hhk}%
:
\bea
\De  m_{K} &=&\left( 3.484\pm 0.006\right) \times 10^{-12}MeV,\hspace{%
1.5cm }\left( \De  m_{K}\right) _{SM}=3.483\times 10^{-12}MeV  \crn
f_{K} &=&160MeV,\hspace{1.5cm}B_{K}=0.85,\hspace{1.5cm}\eta _{K}=0.57,
\hspace{1.5cm}m_{K}=497.614MeV.\nn
\eea
\bea
\left( \De  m_{B_{d}}\right) _{\exp } &=&\left( 3.337\pm 0.033\right)
\times 10^{-10}MeV,\hspace{1.5cm}\left( \De  m_{B_{d}}\right)
_{SM}=3.582\times 10^{-10}MeV,  \crn
f_{B_{d}} &=&188MeV,\hspace{1.5cm}B_{B_{d}}=1.26,\hspace{1.5cm}\eta
_{B_{d}}=0.55,\hspace{1.5cm}m_{B_{d}}=5279.5MeV.\nn
\eea
\bea
\left( \De  m_{B_{s}}\right) _{\exp } &=&\left( 104.19\pm 0.8\right)
\times 10^{-10}MeV,\hspace{1.5cm}\left( \De  m_{B_{s}}\right)
_{SM}=121.103\times 10^{-10}MeV,  \crn
f_{B_{s}} &=&225MeV,\hspace{1.5cm}B_{B_{s}}=1.26,\hspace{1.5cm}\eta
_{B_{s}}=0.55,\hspace{1.5cm}m_{B_{s}}=5366.3MeV.\nn
\eea
\begin{figure}[h]
\includegraphics[width=0.51\textwidth]{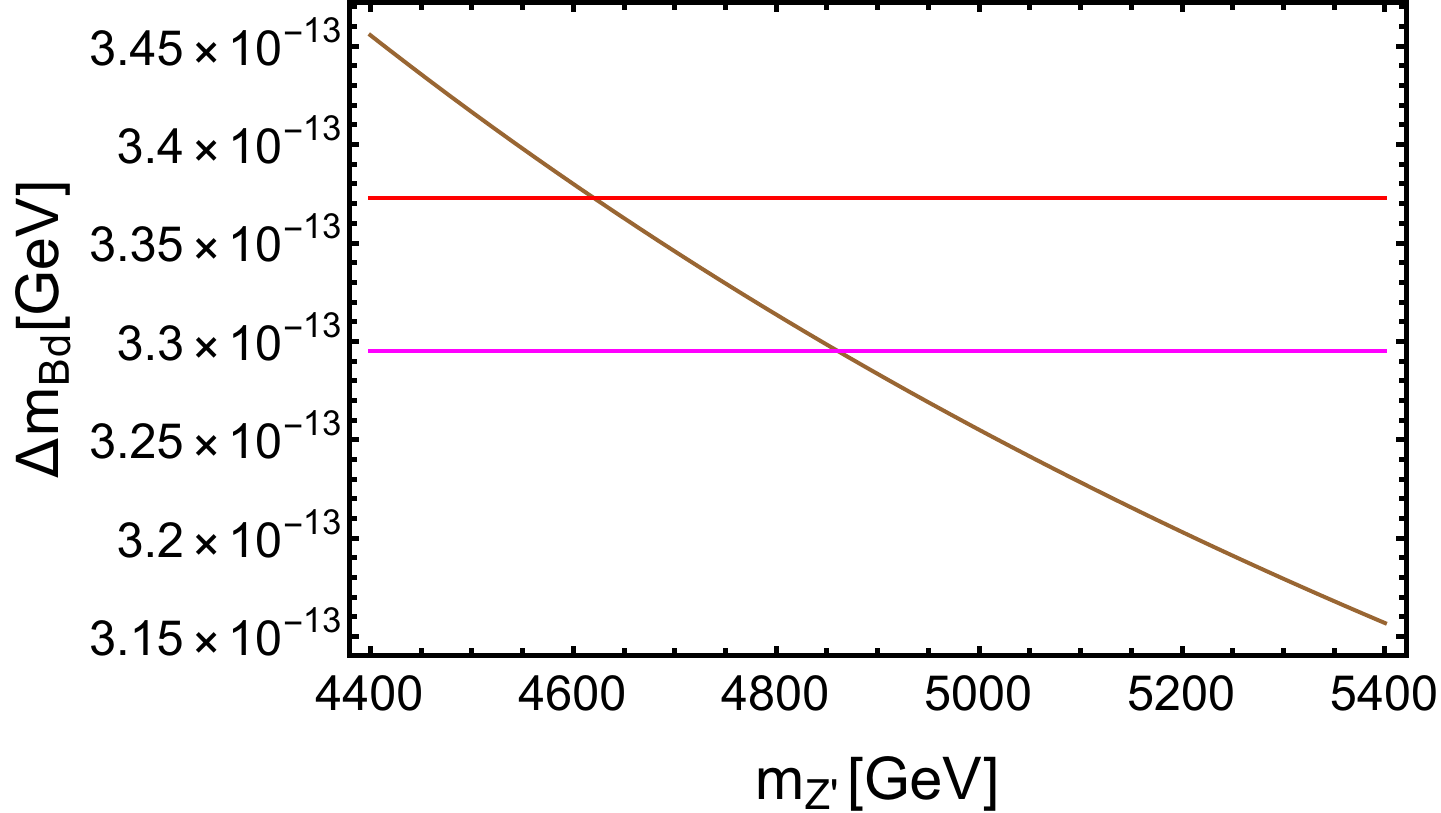}
\caption{The $B_{d}^{0}-\bar{B}_{d}^{0}$ mass splitting as as function of
the $Z^{\prime }$ mass.}
\label{mesons}
\end{figure}
We plot in figure \ref{mesons} the $B_{d}^{0}-\bar{B}_{d}^{0}$ mass
splitting as as function of the $Z^{\prime }$ mass. As seen from figure \ref%
{mesons}, the obtained values for the $B_{s}^{0}-\bar{B}_{s}^{0}$ mass
difference are consistent with the experimental data where the $Z^{\prime }$
mass is larger than about $4.6$ TeV and lower than about $4.9$ TeV. Regarding the $K^{0}-\bar{K}^{0}$, $B_{d}^{0}-\bar{B}_{d}^{0}$ mass
splittings, we have numerically checked that the obtained values are in
accordance with the meson oscillation experimental data in the above
described region of parameter space.

\section{Conclusions}

\label{conclusions} We have constructed a theory based on the $SU(3)_C\times
SU(3)_L\times U(1)_X$ gauge symmetry, where the scalar sector is composed of
two $SU(3)_L$ scalar triplets and several gauge singlet scalar fields. The
theory incorporates the $D_4$ family symmetry, which is supplemented by
several auxiliary cyclic symmetries, whose spontaneous breaking yield viable
and predictive fermion mass matrix textures with hierarchical entries thus
allowing a natural explanation of the current hierarchy of SM fermion masses
and mixings. The tiny masses of the light active neutrinos are produced by
an inverse seesaw mechanism mediated by three right handed Majorana
neutrinos. The smallness of the $\mu$ parameter of the inverse seesaw,
generated after the spontaneous breaking of the discrete symmetries of the
model, is attributed to a right-handed neutrino nonrenormalizable Yukawa
terms. Our proposed model is consistent with Higgs diphoton decay rate
constraints,with the muon anomalous magnetic moment and the meson oscillation experimental data. The consistency of our model with the
muon anomalous magnetic moment requires charged exotic vector like leptons at the TeV scale.

\section*{Acknowledgments}

This research has received funding from ANID-Chile FONDECYT 1210378, ANID PIA/APOYO AFB180002 and \crb{ANID- Programa Milenio - code ICN2019\_044}. \crb{H. N. Long acknowledges the financial support of the International Centre of Physics at the Institute of Physics, VAST under grant No: ICP.2022.02.
}

\appendix

\section{The product rules for $D_{4}$}

\label{app} The $D_4$ group has four singlets and one doublet, $\mathbf{1}%
_{++}$, $\mathbf{1}_{+-}$, $\mathbf{1}_{-+}$, $\mathbf{1}_{--}$, and $%
\mathbf{2}$, respectively. The multiplication of the singlets is simply
given by
\begin{align}
\mathbf{1}_{x_1 y_1} \times \mathbf{1}_{x_2 y_2} = \mathbf{1}_{x_3 y_3}
\end{align}
where $x_3 = x_1 x_2$ and $y_3 = y_1 y_2$. While the tensor product for two
doublets, $\mathbf{a} = (a_1,a_2)^T$ and $\mathbf{b} = (b_1 ,b_2)^T$, is
\begin{align}
\mathbf{a} \times \mathbf{b} = (a_1 b_2 + a_2 b_1)_{\mathbf{1}_{++}} + (a_1
b_2 - a_2 b_1)_{\mathbf{1}_{--}} + (a_1 b_1 + a_2 b_2)_{\mathbf{1}_{+-}} +
(a_1 b_1 - a_2 b_2)_{\mathbf{1}_{-+}}.
\end{align}

\bibliographystyle{utphys}
\bibliography{Refs331D4March30th}

\providecommand{\href}[2]{#2}\begingroup\raggedright\begin{thebibliography}{10}

\bibitem{King:2013eh}
S.~F. King and C.~Luhn, ``{Neutrino Mass and Mixing with Discrete Symmetry},''
  \href{http://dx.doi.org/10.1088/0034-4885/76/5/056201}{{\em Rept. Prog.
  Phys.} {\bfseries 76} (2013) 056201},
  \href{http://arxiv.org/abs/1301.1340}{{\ttfamily arXiv:1301.1340 [hep-ph]}}.

\bibitem{Altarelli:2010gt}
G.~Altarelli and F.~Feruglio, ``{Discrete Flavor Symmetries and Models of
  Neutrino Mixing},'' \href{http://dx.doi.org/10.1103/RevModPhys.82.2701}{{\em
  Rev. Mod. Phys.} {\bfseries 82} (2010) 2701--2729},
  \href{http://arxiv.org/abs/1002.0211}{{\ttfamily arXiv:1002.0211 [hep-ph]}}.

\bibitem{Ishimori:2010au}
H.~Ishimori, T.~Kobayashi, H.~Ohki, Y.~Shimizu, H.~Okada, and M.~Tanimoto,
  ``{Non-Abelian Discrete Symmetries in Particle Physics},''
  \href{http://dx.doi.org/10.1143/PTPS.183.1}{{\em Prog. Theor. Phys. Suppl.}
  {\bfseries 183} (2010) 1--163},
  \href{http://arxiv.org/abs/1003.3552}{{\ttfamily arXiv:1003.3552 [hep-th]}}.

\bibitem{King:2015aea}
S.~F. King, ``{Models of Neutrino Mass, Mixing and CP Violation},''
  \href{http://dx.doi.org/10.1088/0954-3899/42/12/123001}{{\em J. Phys. G}
  {\bfseries 42} (2015) 123001},
  \href{http://arxiv.org/abs/1510.02091}{{\ttfamily arXiv:1510.02091
  [hep-ph]}}.

\bibitem{Frampton:1994rk}
P.~H. Frampton and T.~W. Kephart, ``{Simple nonAbelian finite flavor groups and
  fermion masses},'' \href{http://dx.doi.org/10.1142/S0217751X95002187}{{\em
  Int. J. Mod. Phys. A} {\bfseries 10} (1995) 4689--4704},
  \href{http://arxiv.org/abs/hep-ph/9409330}{{\ttfamily arXiv:hep-ph/9409330}}.

\bibitem{Grimus:2003kq}
W.~Grimus and L.~Lavoura, ``{A Discrete symmetry group for maximal atmospheric
  neutrino mixing},''
  \href{http://dx.doi.org/10.1016/j.physletb.2003.08.032}{{\em Phys. Lett. B}
  {\bfseries 572} (2003) 189--195},
  \href{http://arxiv.org/abs/hep-ph/0305046}{{\ttfamily arXiv:hep-ph/0305046}}.

\bibitem{Grimus:2004rj}
W.~Grimus, A.~S. Joshipura, S.~Kaneko, L.~Lavoura, and M.~Tanimoto, ``{Lepton
  mixing angle $\theta_{13} = 0$ with a horizontal symmetry $D_4$},''
  \href{http://dx.doi.org/10.1088/1126-6708/2004/07/078}{{\em JHEP} {\bfseries
  07} (2004) 078}, \href{http://arxiv.org/abs/hep-ph/0407112}{{\ttfamily
  arXiv:hep-ph/0407112}}.

\bibitem{Frigerio:2004jg}
M.~Frigerio, S.~Kaneko, E.~Ma, and M.~Tanimoto, ``{Quaternion family symmetry
  of quarks and leptons},''
  \href{http://dx.doi.org/10.1103/PhysRevD.71.011901}{{\em Phys. Rev. D}
  {\bfseries 71} (2005) 011901},
  \href{http://arxiv.org/abs/hep-ph/0409187}{{\ttfamily arXiv:hep-ph/0409187}}.

\bibitem{Blum:2007jz}
A.~Blum, C.~Hagedorn, and M.~Lindner, ``{Fermion Masses and Mixings from
  Dihedral Flavor Symmetries with Preserved Subgroups},''
  \href{http://dx.doi.org/10.1103/PhysRevD.77.076004}{{\em Phys. Rev. D}
  {\bfseries 77} (2008) 076004},
  \href{http://arxiv.org/abs/0709.3450}{{\ttfamily arXiv:0709.3450 [hep-ph]}}.

\bibitem{Adulpravitchai:2008yp}
A.~Adulpravitchai, A.~Blum, and C.~Hagedorn, ``{A Supersymmetric D4 Model for
  mu-tau Symmetry},''
  \href{http://dx.doi.org/10.1088/1126-6708/2009/03/046}{{\em JHEP} {\bfseries
  03} (2009) 046}, \href{http://arxiv.org/abs/0812.3799}{{\ttfamily
  arXiv:0812.3799 [hep-ph]}}.

\bibitem{Ishimori:2008gp}
H.~Ishimori, T.~Kobayashi, H.~Ohki, Y.~Omura, R.~Takahashi, and M.~Tanimoto,
  ``{D(4) Flavor Symmetry for Neutrino Masses and Mixing},''
  \href{http://dx.doi.org/10.1016/j.physletb.2008.03.007}{{\em Phys. Lett. B}
  {\bfseries 662} (2008) 178--184},
  \href{http://arxiv.org/abs/0802.2310}{{\ttfamily arXiv:0802.2310 [hep-ph]}}.

\bibitem{Hagedorn:2010mq}
C.~Hagedorn and R.~Ziegler, ``{$\mu-\tau$ Symmetry and Charged Lepton Mass
  Hierarchy in a Supersymmetric $D_4$ Model},''
  \href{http://dx.doi.org/10.1103/PhysRevD.82.053011}{{\em Phys. Rev. D}
  {\bfseries 82} (2010) 053011},
  \href{http://arxiv.org/abs/1007.1888}{{\ttfamily arXiv:1007.1888 [hep-ph]}}.

\bibitem{Meloni:2011cc}
D.~Meloni, S.~Morisi, and E.~Peinado, ``{Stability of dark matter from the
  D4xZ2 flavor group},''
  \href{http://dx.doi.org/10.1016/j.physletb.2011.07.084}{{\em Phys. Lett. B}
  {\bfseries 703} (2011) 281--287},
  \href{http://arxiv.org/abs/1104.0178}{{\ttfamily arXiv:1104.0178 [hep-ph]}}.

\bibitem{Vien:2013zra}
V.~V. Vien and H.~N. Long, ``{The $D_4$ flavor symmery in 3-3-1 model with
  neutral leptons},'' \href{http://dx.doi.org/10.1142/S0217751X13501595}{{\em
  Int. J. Mod. Phys. A} {\bfseries 28} (2013) 1350159},
  \href{http://arxiv.org/abs/1312.5034}{{\ttfamily arXiv:1312.5034 [hep-ph]}}.

\bibitem{Vien:2014ica}
V.~V. Vien and H.~N. Long, ``{Quark masses and mixings in the 3-3-1 model with
  neutral leptons based on $D_{4}$ flavor symmetry},''
  \href{http://dx.doi.org/10.3938/jkps.66.1809}{{\em J. Korean Phys. Soc.}
  {\bfseries 66} no.~12, (2015) 1809--1815},
  \href{http://arxiv.org/abs/1408.4333}{{\ttfamily arXiv:1408.4333 [hep-ph]}}.

\bibitem{Vien:2014soa}
V.~V. Vien, ``{Neutrino mass and mixing in the 3-3-1 model with neutral leptons
  based on D4 flavor symmetry},''
  \href{http://dx.doi.org/10.1142/S0217732314501223}{{\em Mod. Phys. Lett. A}
  {\bfseries 29} (2014) 1450122}.

\bibitem{CarcamoHernandez:2020ney}
A.~E. C\'arcamo~Hern\'andez, C.~O. Dib, and U.~J. Salda\~na Salazar, ``{When
  $\tan \beta$ meets all the mixing angles},''
  \href{http://dx.doi.org/10.1016/j.physletb.2020.135750}{{\em Phys. Lett. B}
  {\bfseries 809} (2020) 135750},
  \href{http://arxiv.org/abs/2001.07140}{{\ttfamily arXiv:2001.07140
  [hep-ph]}}.

\bibitem{Vien:2020uzf}
V.~V. Vien, ``{Fermion mass and mixing in the $U(1)_{B-L}$ extension of the
  standard model with $D_4$ symmetry},''
  \href{http://dx.doi.org/10.1088/1361-6471/ab7ec0}{{\em J. Phys. G} {\bfseries
  47} no.~5, (2020) 055007}.

\bibitem{Vien:2021diw}
V.~V. Vien, ``{Fermion mass hierarchies and mixings in a $B-L$ model with
  $D_4\times Z_4\times Z_2$ symmetry},''
  \href{http://arxiv.org/abs/2111.14701}{{\ttfamily arXiv:2111.14701
  [hep-ph]}}.

\bibitem{Bonilla:2020hct}
C.~Bonilla, L.~M.~G. de~la Vega, R.~Ferro-Hernandez, N.~Nath, and E.~Peinado,
  ``{Neutrino phenomenology in a left-right $D_4$ symmetric model},''
  \href{http://dx.doi.org/10.1103/PhysRevD.102.036006}{{\em Phys. Rev. D}
  {\bfseries 102} no.~3, (2020) 036006},
  \href{http://arxiv.org/abs/2003.06444}{{\ttfamily arXiv:2003.06444
  [hep-ph]}}.

\bibitem{Athron:2021iuf}
P.~Athron, C.~Bal\'azs, D.~H.~J. Jacob, W.~Kotlarski, D.~St\"ockinger, and
  H.~St\"ockinger-Kim, ``{New physics explanations of $a_{\mu}$ in light of the
  FNAL muon g-2 measurement},''
  \href{http://dx.doi.org/10.1007/JHEP09(2021)080}{{\em JHEP} {\bfseries 09}
  (2021) 080}, \href{http://arxiv.org/abs/2104.03691}{{\ttfamily
  arXiv:2104.03691 [hep-ph]}}.

\bibitem{Valle:1983dk}
J.~W.~F. Valle and M.~Singer, ``{Lepton Number Violation With Quasi Dirac
  Neutrinos},''
\href{http://dx.doi.org/10.1103/PhysRevD.28.540}{{\em Phys. Rev.} {\bfseries
  D28} (1983) 540}.

\bibitem{Pisano:1991ee}
F.~Pisano and V.~Pleitez, ``{An SU(3) x U(1) model for electroweak
  interactions},'' \href{http://dx.doi.org/10.1103/PhysRevD.46.410}{{\em Phys.
  Rev.} {\bfseries D46} (1992) 410--417},
\href{http://arxiv.org/abs/hep-ph/9206242}{{\ttfamily arXiv:hep-ph/9206242
  [hep-ph]}}.

\bibitem{Frampton:1992wt}
P.~H. Frampton, ``{Chiral dilepton model and the flavor question},''
\href{http://dx.doi.org/10.1103/PhysRevLett.69.2889}{{\em Phys. Rev. Lett.}
  {\bfseries 69} (1992) 2889--2891}.

\bibitem{Foot:1994ym}
R.~Foot, H.~N. Long, and T.~A. Tran, ``{$SU(3)_L \otimes U(1)_N$ and $SU(4)_L
  \otimes U(1)_N$ gauge models with right-handed neutrinos},''
  \href{http://dx.doi.org/10.1103/PhysRevD.50.R34}{{\em Phys. Rev. D}
  {\bfseries 50} no.~1, (1994) R34--R38},
  \href{http://arxiv.org/abs/hep-ph/9402243}{{\ttfamily arXiv:hep-ph/9402243}}.

\bibitem{Hoang:1995vq}
H.~N. Long, ``{The 331 model with right handed neutrinos},''
  \href{http://dx.doi.org/10.1103/PhysRevD.53.437}{{\em Phys. Rev.} {\bfseries
  D53} (1996) 437--445},
\href{http://arxiv.org/abs/hep-ph/9504274}{{\ttfamily arXiv:hep-ph/9504274
  [hep-ph]}}.

\bibitem{CarcamoHernandez:2005ka}
A.~E. Carcamo~Hernandez, R.~Martinez, and F.~Ochoa, ``{Z and Z' decays with and
  without FCNC in 331 models},''
  \href{http://dx.doi.org/10.1103/PhysRevD.73.035007}{{\em Phys. Rev.}
  {\bfseries D73} (2006) 035007},
\href{http://arxiv.org/abs/hep-ph/0510421}{{\ttfamily arXiv:hep-ph/0510421
  [hep-ph]}}.

\bibitem{Chang:2006aa}
D.~Chang and H.~N. Long, ``{Interesting radiative patterns of neutrino mass in
  an SU(3)(C) x SU(3)(L) x U(1)(X) model with right-handed neutrinos},''
  \href{http://dx.doi.org/10.1103/PhysRevD.73.053006}{{\em Phys. Rev.}
  {\bfseries D73} (2006) 053006},
\href{http://arxiv.org/abs/hep-ph/0603098}{{\ttfamily arXiv:hep-ph/0603098
  [hep-ph]}}.

\bibitem{Hernandez:2013mcf}
A.~E. Carcamo~Hernandez, R.~Martinez, and F.~Ochoa, ``{Radiative seesaw-type
  mechanism of quark masses in $SU(3)_C \otimes SU(3)_L \otimes U(1)_X$},''
  \href{http://dx.doi.org/10.1103/PhysRevD.87.075009}{{\em Phys. Rev.}
  {\bfseries D87} no.~7, (2013) 075009},
\href{http://arxiv.org/abs/1302.1757}{{\ttfamily arXiv:1302.1757 [hep-ph]}}.

\bibitem{Hernandez:2013hea}
A.~E. Cárcamo~Hernández, R.~Martinez, and F.~Ochoa, ``{Fermion masses and
  mixings in the 3-3-1 model with right-handed neutrinos based on the $S_3$
  flavor symmetry},''
  \href{http://dx.doi.org/10.1140/epjc/s10052-016-4480-3}{{\em Eur. Phys. J.}
  {\bfseries C76} no.~11, (2016) 634},
\href{http://arxiv.org/abs/1309.6567}{{\ttfamily arXiv:1309.6567 [hep-ph]}}.

\bibitem{Boucenna:2014ela}
S.~M. Boucenna, S.~Morisi, and J.~W.~F. Valle, ``{Radiative neutrino mass in
  3-3-1 scheme},'' \href{http://dx.doi.org/10.1103/PhysRevD.90.013005}{{\em
  Phys. Rev.} {\bfseries D90} no.~1, (2014) 013005},
\href{http://arxiv.org/abs/1405.2332}{{\ttfamily arXiv:1405.2332 [hep-ph]}}.

\bibitem{Hernandez:2014lpa}
A.~E. Cárcamo~Hernández, E.~Cataño~Mur, and R.~Martinez, ``{Lepton masses
  and mixing in $SU(3)_{C}\otimes SU(3)_{L}\otimes U(1)_{X}$ models with a
  $S_3$ flavor symmetry},''
  \href{http://dx.doi.org/10.1103/PhysRevD.90.073001}{{\em Phys. Rev.}
  {\bfseries D90} no.~7, (2014) 073001},
\href{http://arxiv.org/abs/1407.5217}{{\ttfamily arXiv:1407.5217 [hep-ph]}}.

\bibitem{Hernandez:2014vta}
A.~E. Cárcamo~Hernández, R.~Martinez, and J.~Nisperuza, ``{$S_3$ discrete
  group as a source of the quark mass and mixing pattern in $331$ models},''
  \href{http://dx.doi.org/10.1140/epjc/s10052-015-3278-z}{{\em Eur. Phys. J.}
  {\bfseries C75} no.~2, (2015) 72},
\href{http://arxiv.org/abs/1401.0937}{{\ttfamily arXiv:1401.0937 [hep-ph]}}.

\bibitem{Okada:2015bxa}
H.~Okada, N.~Okada, and Y.~Orikasa, ``{Radiative seesaw mechanism in a minimal
  3-3-1 model},'' \href{http://dx.doi.org/10.1103/PhysRevD.93.073006}{{\em
  Phys. Rev.} {\bfseries D93} no.~7, (2016) 073006},
\href{http://arxiv.org/abs/1504.01204}{{\ttfamily arXiv:1504.01204 [hep-ph]}}.

\bibitem{Hernandez:2016eod}
A.~E. Cárcamo~Hernández, H.~N. Long, and V.~V. Vien, ``{A 3-3-1 model with
  right-handed neutrinos based on the $\varDelta \left( 27\right) $ family
  symmetry},'' \href{http://dx.doi.org/10.1140/epjc/s10052-016-4074-0}{{\em
  Eur. Phys. J.} {\bfseries C76} no.~5, (2016) 242},
\href{http://arxiv.org/abs/1601.05062}{{\ttfamily arXiv:1601.05062 [hep-ph]}}.

\bibitem{Fonseca:2016tbn}
R.~M. Fonseca and M.~Hirsch, ``{A flipped 331 model},''
  \href{http://dx.doi.org/10.1007/JHEP08(2016)003}{{\em JHEP} {\bfseries 08}
  (2016) 003},
\href{http://arxiv.org/abs/1606.01109}{{\ttfamily arXiv:1606.01109 [hep-ph]}}.

\bibitem{CarcamoHernandez:2017cwi}
A.~E. Cárcamo~Hernández, S.~Kovalenko, H.~N. Long, and I.~Schmidt, ``{A
  variant of 3-3-1 model for the generation of the SM fermion mass and mixing
  pattern},'' \href{http://dx.doi.org/10.1007/JHEP07(2018)144}{{\em JHEP}
  {\bfseries 07} (2018) 144},
\href{http://arxiv.org/abs/1705.09169}{{\ttfamily arXiv:1705.09169 [hep-ph]}}.

\bibitem{CarcamoHernandez:2018iel}
A.~E. Cárcamo~Hernández, H.~N. Long, and V.~V. Vien, ``{The first
  $\Delta(27)$ flavor 3-3-1 model with low scale seesaw mechanism},''
  \href{http://dx.doi.org/10.1140/epjc/s10052-018-6284-0}{{\em Eur. Phys. J.}
  {\bfseries C78} no.~10, (2018) 804},
\href{http://arxiv.org/abs/1803.01636}{{\ttfamily arXiv:1803.01636 [hep-ph]}}.

\bibitem{CarcamoHernandez:2019vih}
A.~E. Cárcamo~Hernández, Y.~Hidalgo~Velásquez, and N.~A. Pérez-Julve, ``{A
  3-3-1 model with low scale seesaw mechanisms},''
  \href{http://dx.doi.org/10.1140/epjc/s10052-019-7325-z}{{\em Eur. Phys. J.}
  {\bfseries C79} no.~10, (2019) 828},
\href{http://arxiv.org/abs/1905.02323}{{\ttfamily arXiv:1905.02323 [hep-ph]}}.

\bibitem{CarcamoHernandez:2019iwh}
A.~E. Cárcamo~Hernández, N.~A. Pérez-Julve, and Y.~Hidalgo~Velásquez,
  ``{Fermion masses and mixings and some phenomenological aspects of a 3-3-1
  model with linear seesaw mechanism},''
  \href{http://dx.doi.org/10.1103/PhysRevD.100.095025}{{\em Phys. Rev.}
  {\bfseries D100} no.~9, (2019) 095025},
\href{http://arxiv.org/abs/1907.13083}{{\ttfamily arXiv:1907.13083 [hep-ph]}}.

\bibitem{CarcamoHernandez:2019lhv}
A.~E. Cárcamo~Hernández, D.~T. Huong, and H.~N. Long, ``{Minimal model for
  the fermion flavor structure, mass hierarchy, dark matter, leptogenesis, and
  the electron and muon anomalous magnetic moments},''
  \href{http://dx.doi.org/10.1103/PhysRevD.102.055002}{{\em Phys. Rev.}
  {\bfseries D102} no.~5, (2020) 055002},
\href{http://arxiv.org/abs/1910.12877}{{\ttfamily arXiv:1910.12877 [hep-ph]}}.

\bibitem{CarcamoHernandez:2020pxw}
A.~E. Cárcamo~Hernández, Y.~Hidalgo~Velásquez, S.~Kovalenko, H.~N. Long,
  N.~A. Pérez-Julve, and V.~V. Vien, ``{Fermion spectrum and $g-2$ anomalies
  in a low scale 3-3-1 model},''
  \href{http://dx.doi.org/10.1140/epjc/s10052-021-08974-4.,
  10.1140/epjc/s10052-021-08974-4}{{\em Eur. Phys. J.} {\bfseries C81} no.~2,
  (2021) 191},
\href{http://arxiv.org/abs/2002.07347}{{\ttfamily arXiv:2002.07347 [hep-ph]}}.

\bibitem{CarcamoHernandez:2020ehn}
A.~E. Cárcamo~Hernández, J.~W.~F. Valle, and C.~A. Vaquera-Araujo, ``{Simple
  theory for scotogenic dark matter with residual matter-parity},''
  \href{http://dx.doi.org/10.1016/j.physletb.2020.135757}{{\em Phys. Lett.}
  {\bfseries B809} (2020) 135757},
\href{http://arxiv.org/abs/2006.06009}{{\ttfamily arXiv:2006.06009 [hep-ph]}}.

\bibitem{Xing:2019vks}
Z.-z. Xing, ``{Flavor structures of charged fermions and massive neutrinos},''
  \href{http://dx.doi.org/10.1016/j.physrep.2020.02.001}{{\em Phys. Rept.}
  {\bfseries 854} (2020) 1--147},
  \href{http://arxiv.org/abs/1909.09610}{{\ttfamily arXiv:1909.09610
  [hep-ph]}}.

\bibitem{Zyla:2020zbs}
{\bfseries Particle Data Group} Collaboration, P.~A. Zyla {\em et~al.},
  ``{Review of Particle Physics},''
  \href{http://dx.doi.org/10.1093/ptep/ptaa104}{{\em PTEP} {\bfseries 2020}
  no.~8, (2020) 083C01}.

\bibitem{Esteban:2020cvm}
I.~Esteban, M.~C. Gonzalez-Garcia, M.~Maltoni, T.~Schwetz, and A.~Zhou, ``{The
  fate of hints: updated global analysis of three-flavor neutrino
  oscillations},'' \href{http://dx.doi.org/10.1007/JHEP09(2020)178}{{\em JHEP}
  {\bfseries 09} (2020) 178}, \href{http://arxiv.org/abs/2007.14792}{{\ttfamily
  arXiv:2007.14792 [hep-ph]}}.

\bibitem{RoyChoudhury:2019hls}
S.~Roy~Choudhury and S.~Hannestad, ``{Updated results on neutrino mass and mass
  hierarchy from cosmology with Planck 2018 likelihoods},''
  \href{http://dx.doi.org/10.1088/1475-7516/2020/07/037}{{\em JCAP} {\bfseries
  2007} (2020) 037},
\href{http://arxiv.org/abs/1907.12598}{{\ttfamily arXiv:1907.12598
  [astro-ph.CO]}}.

\bibitem{Shifman:1979eb}
M.~A. Shifman, A.~I. Vainshtein, M.~B. Voloshin, and V.~I. Zakharov,
  ``{Low-Energy Theorems for Higgs Boson Couplings to Photons},'' {\em Sov. J.
  Nucl. Phys.} {\bfseries 30} (1979) 711--716.

\bibitem{Gavela:1981ri}
M.~B. Gavela, G.~Girardi, C.~Malleville, and P.~Sorba, ``{A Nonlinear R(xi)
  Gauge Condition for the Electroweak SU(2) X U(1) Model},''
  \href{http://dx.doi.org/10.1016/0550-3213(81)90529-0}{{\em Nucl. Phys. B}
  {\bfseries 193} (1981) 257--268}.

\bibitem{Kalyniak:1985ct}
P.~Kalyniak, R.~Bates, and J.~N. Ng, ``{Two Photon Decays of Scalar and
  Pseudoscalar Bosons in Supersymmetry},''
  \href{http://dx.doi.org/10.1103/PhysRevD.33.755}{{\em Phys. Rev. D}
  {\bfseries 33} (1986) 755}.

\bibitem{Spira:1997dg}
M.~Spira, ``{QCD effects in Higgs physics},''
  \href{http://dx.doi.org/10.1002/(SICI)1521-3978(199804)46:3<203::AID-PROP203>3.0.CO;2-4}{{\em
  Fortsch. Phys.} {\bfseries 46} (1998) 203--284},
  \href{http://arxiv.org/abs/hep-ph/9705337}{{\ttfamily arXiv:hep-ph/9705337}}.

\bibitem{Djouadi:2005gj}
A.~Djouadi, ``{The Anatomy of electro-weak symmetry breaking. II. The Higgs
  bosons in the minimal supersymmetric model},''
  \href{http://dx.doi.org/10.1016/j.physrep.2007.10.005}{{\em Phys. Rept.}
  {\bfseries 459} (2008) 1--241},
  \href{http://arxiv.org/abs/hep-ph/0503173}{{\ttfamily arXiv:hep-ph/0503173}}.

\bibitem{Marciano:2011gm}
W.~J. Marciano, C.~Zhang, and S.~Willenbrock, ``{Higgs Decay to Two Photons},''
  \href{http://dx.doi.org/10.1103/PhysRevD.85.013002}{{\em Phys. Rev. D}
  {\bfseries 85} (2012) 013002},
  \href{http://arxiv.org/abs/1109.5304}{{\ttfamily arXiv:1109.5304 [hep-ph]}}.

\bibitem{Wang:2012gm}
L.~Wang and X.-F. Han, ``{The recent Higgs boson data and Higgs triplet model
  with vector-like quark},''
  \href{http://dx.doi.org/10.1103/PhysRevD.86.095007}{{\em Phys. Rev. D}
  {\bfseries 86} (2012) 095007},
  \href{http://arxiv.org/abs/1206.1673}{{\ttfamily arXiv:1206.1673 [hep-ph]}}.

\bibitem{Sirunyan:2018ouh}
{\bfseries CMS} Collaboration, A.~M. Sirunyan {\em et~al.}, ``{Measurements of
  Higgs boson properties in the diphoton decay channel in proton-proton
  collisions at $\sqrt{s} =$ 13 TeV},''
  \href{http://dx.doi.org/10.1007/JHEP11(2018)185}{{\em JHEP} {\bfseries 11}
  (2018) 185}, \href{http://arxiv.org/abs/1804.02716}{{\ttfamily
  arXiv:1804.02716 [hep-ex]}}.

\bibitem{Aad:2019mbh}
{\bfseries ATLAS} Collaboration, G.~Aad {\em et~al.}, ``{Combined measurements
  of Higgs boson production and decay using up to $80$ fb$^{-1}$ of
  proton-proton collision data at $\sqrt{s}=$ 13 TeV collected with the ATLAS
  experiment},'' \href{http://dx.doi.org/10.1103/PhysRevD.101.012002}{{\em
  Phys. Rev. D} {\bfseries 101} no.~1, (2020) 012002},
  \href{http://arxiv.org/abs/1909.02845}{{\ttfamily arXiv:1909.02845
  [hep-ex]}}.

\bibitem{Diaz:2002uk}
R.~A. Diaz, R.~Martinez, and J.~A. Rodriguez, ``{Phenomenology of lepton flavor
  violation in 2HDM(3) from (g-2)(mu) and leptonic decays},''
  \href{http://dx.doi.org/10.1103/PhysRevD.67.075011}{{\em Phys. Rev.}
  {\bfseries D67} (2003) 075011},
\href{http://arxiv.org/abs/hep-ph/0208117}{{\ttfamily arXiv:hep-ph/0208117
  [hep-ph]}}.

\bibitem{Jegerlehner:2009ry}
F.~Jegerlehner and A.~Nyffeler, ``{The Muon g-2},''
  \href{http://dx.doi.org/10.1016/j.physrep.2009.04.003}{{\em Phys. Rept.}
  {\bfseries 477} (2009) 1--110},
\href{http://arxiv.org/abs/0902.3360}{{\ttfamily arXiv:0902.3360 [hep-ph]}}.

\bibitem{Kelso:2014qka}
C.~Kelso, H.~N. Long, R.~Martinez, and F.~S. Queiroz, ``{Connection of
  $g-2_{\mu}$, electroweak, dark matter, and collider constraints on 331
  models},'' \href{http://dx.doi.org/10.1103/PhysRevD.90.113011}{{\em Phys.
  Rev.} {\bfseries D90} no.~11, (2014) 113011},
\href{http://arxiv.org/abs/1408.6203}{{\ttfamily arXiv:1408.6203 [hep-ph]}}.

\bibitem{Lindner:2016bgg}
M.~Lindner, M.~Platscher, and F.~S. Queiroz, ``{A Call for New Physics : The
  Muon Anomalous Magnetic Moment and Lepton Flavor Violation},''
  \href{http://dx.doi.org/10.1016/j.physrep.2017.12.001}{{\em Phys. Rept.}
  {\bfseries 731} (2018) 1--82},
\href{http://arxiv.org/abs/1610.06587}{{\ttfamily arXiv:1610.06587 [hep-ph]}}.

\bibitem{Kowalska:2017iqv}
K.~Kowalska and E.~M. Sessolo, ``{Expectations for the muon g-2 in simplified
  models with dark matter},''
  \href{http://dx.doi.org/10.1007/JHEP09(2017)112}{{\em JHEP} {\bfseries 09}
  (2017) 112}, \href{http://arxiv.org/abs/1707.00753}{{\ttfamily
  arXiv:1707.00753 [hep-ph]}}.

\bibitem{Hagiwara:2011af}
K.~Hagiwara, R.~Liao, A.~D. Martin, D.~Nomura, and T.~Teubner, ``{$(g-2)_\mu$
  and $\alpha(M^2_Z)$ re-evaluated using new precise data},''
  \href{http://dx.doi.org/10.1088/0954-3899/38/8/085003}{{\em J. Phys.}
  {\bfseries G38} (2011) 085003},
\href{http://arxiv.org/abs/1105.3149}{{\ttfamily arXiv:1105.3149 [hep-ph]}}.

\bibitem{Davier:2017zfy}
M.~Davier, A.~Hoecker, B.~Malaescu, and Z.~Zhang, ``{Reevaluation of the
  hadronic vacuum polarisation contributions to the Standard Model predictions
  of the muon $g-2$ and ${\alpha (m_Z^2)}$ using newest hadronic cross-section
  data},'' \href{http://dx.doi.org/10.1140/epjc/s10052-017-5161-6}{{\em Eur.
  Phys. J.} {\bfseries C77} no.~12, (2017) 827},
\href{http://arxiv.org/abs/1706.09436}{{\ttfamily arXiv:1706.09436 [hep-ph]}}.

\bibitem{Blum:2018mom}
{\bfseries RBC, UKQCD} Collaboration, T.~Blum, P.~A. Boyle, V.~Gülpers,
  T.~Izubuchi, L.~Jin, C.~Jung, A.~Jüttner, C.~Lehner, A.~Portelli, and J.~T.
  Tsang, ``{Calculation of the hadronic vacuum polarization contribution to the
  muon anomalous magnetic moment},''
  \href{http://dx.doi.org/10.1103/PhysRevLett.121.022003}{{\em Phys. Rev.
  Lett.} {\bfseries 121} no.~2, (2018) 022003},
\href{http://arxiv.org/abs/1801.07224}{{\ttfamily arXiv:1801.07224 [hep-lat]}}.

\bibitem{Keshavarzi:2018mgv}
A.~Keshavarzi, D.~Nomura, and T.~Teubner, ``{Muon $g-2$ and $\alpha(M_Z^2)$: a
  new data-based analysis},''
  \href{http://dx.doi.org/10.1103/PhysRevD.97.114025}{{\em Phys. Rev.}
  {\bfseries D97} no.~11, (2018) 114025},
\href{http://arxiv.org/abs/1802.02995}{{\ttfamily arXiv:1802.02995 [hep-ph]}}.

\bibitem{Nomura:2018lsx}
T.~Nomura and H.~Okada, ``{One-loop neutrino mass model without any additional
  symmetries},'' \href{http://dx.doi.org/10.1016/j.dark.2019.100359}{{\em Phys.
  Dark Univ.} {\bfseries 26} (2019) 100359},
\href{http://arxiv.org/abs/1808.05476}{{\ttfamily arXiv:1808.05476 [hep-ph]}}.

\bibitem{Nomura:2018vfz}
T.~Nomura and H.~Okada, ``{Zee-Babu type model with $U(1)_{L_\mu - L_\tau}$
  gauge symmetry},'' \href{http://dx.doi.org/10.1103/PhysRevD.97.095023}{{\em
  Phys. Rev.} {\bfseries D97} no.~9, (2018) 095023},
\href{http://arxiv.org/abs/1803.04795}{{\ttfamily arXiv:1803.04795 [hep-ph]}}.

\bibitem{Aoyama:2020ynm}
T.~Aoyama {\em et~al.}, ``{The anomalous magnetic moment of the muon in the
  Standard Model},''
  \href{http://dx.doi.org/10.1016/j.physrep.2020.07.006}{{\em Phys. Rept.}
  {\bfseries 887} (2020) 1--166},
\href{http://arxiv.org/abs/2006.04822}{{\ttfamily arXiv:2006.04822 [hep-ph]}}.

\bibitem{Abi:2021gix}
{\bfseries Muon g-2} Collaboration, B.~Abi {\em et~al.}, ``{Measurement of the
  Positive Muon Anomalous Magnetic Moment to 0.46 ppm},''
  \href{http://dx.doi.org/10.1103/PhysRevLett.126.141801}{{\em Phys. Rev.
  Lett.} {\bfseries 126} no.~14, (2021) 141801},
\href{http://arxiv.org/abs/2104.03281}{{\ttfamily arXiv:2104.03281 [hep-ex]}}.

\bibitem{Queiroz:2016gif}
F.~S. Queiroz, C.~Siqueira, and J.~W.~F. Valle, ``{Constraining Flavor Changing
  Interactions from LHC Run-2 Dilepton Bounds with Vector Mediators},''
  \href{http://dx.doi.org/10.1016/j.physletb.2016.10.057}{{\em Phys. Lett. B}
  {\bfseries 763} (2016) 269--274},
  \href{http://arxiv.org/abs/1608.07295}{{\ttfamily arXiv:1608.07295
  [hep-ph]}}.

\bibitem{Dedes:2002er}
A.~Dedes and A.~Pilaftsis, ``{Resummed Effective Lagrangian for Higgs Mediated
  FCNC Interactions in the CP Violating MSSM},''
  \href{http://dx.doi.org/10.1103/PhysRevD.67.015012}{{\em Phys. Rev. D}
  {\bfseries 67} (2003) 015012},
  \href{http://arxiv.org/abs/hep-ph/0209306}{{\ttfamily arXiv:hep-ph/0209306}}.

\bibitem{Aranda:2012bv}
A.~Aranda, C.~Bonilla, and J.~L. Diaz-Cruz, ``{Three generations of Higgses and
  the cyclic groups},''
  \href{http://dx.doi.org/10.1016/j.physletb.2012.09.011}{{\em Phys. Lett. B}
  {\bfseries 717} (2012) 248--251},
  \href{http://arxiv.org/abs/1204.5558}{{\ttfamily arXiv:1204.5558 [hep-ph]}}.

\bibitem{Khalil:2013ixa}
S.~Khalil and S.~Salem, ``{Enhancement of $H \to \gamma\gamma$ in $SU(5)$ model
  with 45$_{H^1}$ plet},''
  \href{http://dx.doi.org/10.1016/j.nuclphysb.2013.08.016}{{\em Nucl. Phys. B}
  {\bfseries 876} (2013) 473--492},
  \href{http://arxiv.org/abs/1304.3689}{{\ttfamily arXiv:1304.3689 [hep-ph]}}.

\bibitem{Buras:2016dxz}
A.~J. Buras and F.~De~Fazio, ``{331 Models Facing the Tensions in $\Delta F=2$
  Processes with the Impact on $\varepsilon^\prime/\varepsilon$,
  $B_s\to\mu^+\mu^-$ and $B\to K^*\mu^+\mu^-$},''
  \href{http://dx.doi.org/10.1007/JHEP08(2016)115}{{\em JHEP} {\bfseries 08}
  (2016) 115}, \href{http://arxiv.org/abs/1604.02344}{{\ttfamily
  arXiv:1604.02344 [hep-ph]}}.

\bibitem{Ferreira:2017tvy}
P.~M. Ferreira, I.~P. Ivanov, E.~Jim\'enez, R.~Pasechnik, and H.~Ser\^odio,
  ``{CP4 miracle: shaping Yukawa sector with CP symmetry of order four},''
  \href{http://dx.doi.org/10.1007/JHEP01(2018)065}{{\em JHEP} {\bfseries 01}
  (2018) 065}, \href{http://arxiv.org/abs/1711.02042}{{\ttfamily
  arXiv:1711.02042 [hep-ph]}}.

\bibitem{Duy:2020hhk}
N.~T. Duy, T.~Inami, and D.~T. Huong, ``{Physical constraints derived from FCNC
  in the 3-3-1-1 model},''
  \href{http://dx.doi.org/https://doi.org/10.1140/epjc/s10052-021-09583-x}{{\em
  Eur. Phys. J. C} {\bfseries 81} (2021) 813},
  \href{http://arxiv.org/abs/2009.09698}{{\ttfamily arXiv:2009.09698
  [hep-ph]}}.

\end{thebibliography}\endgroup


\end{document}